\documentclass[11pt]{article}
\usepackage{amsmath,amssymb}

\makeatletter
\@addtoreset{equation}{section}
\makeatother

\def\slashchar#1{\setbox0=\hbox{$#1$}           
   \dimen0=\wd0                                 
   \setbox1=\hbox{/} \dimen1=\wd1               
   \ifdim\dimen0>\dimen1                        
      \rlap{\hbox to \dimen0{\hfil/\hfil}}      
      #1                                        
   \else                                        
      \rlap{\hbox to \dimen1{\hfil$#1$\hfil}}   
      /                                         
   \fi}

\def\writecenter#1{
   \rlap{\hbox to 50mm{\hfil#1\hfil}}   
   }

\def\nn{\nonumber}

\def\be{\begin{equation}}
\def\ee{\end{equation}}
\def\ben{\begin{displaymath}}
\def\een{\end{displaymath}}
\def\bea{\begin{eqnarray}}
\def\eea{\end{eqnarray}}
\def\ft#1#2{{\textstyle {\frac{#1}{#2}} }}

\makeatletter \@addtoreset{equation}{section} \makeatother

\def\cE{{\cal E}}
\def\cF{{\cal F}}
\def\cG{{\cal G}}

\def\cL{{\cal L}}

\def\c{\gamma}
\def\C{\Gamma}

\def\eps{\epsilon}

\def\x2{X}

\newcommand{\w}[1]{\\[0.#1cm]}

\def\eq#1{(\ref{#1})}
\def\ft#1#2{{\textstyle{{\scriptstyle #1}\over {\scriptstyle #2}}}}

\def\s2{{\sqrt 2}}

\def\tf{{\widetilde f}}

\def\ptau{{\partial\over\partial\tau}}
\def\pphi{{\partial\over\partial\phi}}
\def\prho{{\partial\over\partial\rho}}

\def\slashchar#1{\setbox0=\hbox{$#1$}           
   \dimen0=\wd0                                 
   \setbox1=\hbox{/} \dimen1=\wd1               
   \ifdim\dimen0>\dimen1                        
      \rlap{\hbox to \dimen0{\hfil/\hfil}}      
      #1                                        
   \else                                        
      \rlap{\hbox to \dimen1{\hfil$#1$\hfil}}   
      /                                         
   \fi}

\newcommand{\hoch}[1]{$\, ^{#1}$}

\newcommand{\tamphys}{\it George and Cynthia Woods Mitchell  Institute
for Fundamental Physics and Astronomy,\\
Texas A\&M University, College Station, TX 77843, USA}

\newcommand{\auth}{
Ergin Sezgin\hoch{\dagger} and Yoshiaki Tanii \hoch{\ddagger}
}

\begin{document}

\begin{flushright}
\hfill{
MIFP-09-13}\\
STUPP-09-201\\

\end{flushright}

\vspace{25pt}

\begin{center}

{\large {\large\bf Witten-Nester Energy in Topologically Massive Gravity}}

\vspace{25pt}
\auth

\vspace{10pt}
\hoch{\dagger}{\tamphys}

\vspace{20pt}

\hoch{\ddagger}{\it  Division of Material Science,
Graduate School of Science and Engineering, \\
Saitama University, Saitama 338-8570, Japan
}

\vspace{40pt}

\underline{ABSTRACT}
\end{center}

We formulate topologically massive supergravity with cosmological constant in the first order formalism, and construct the Noether supercurrent and superpotential associated with its local supersymmetry. Using these results, we construct in ordinary topologically massive gravity the Witten-Nester integral for conserved charges containing  spinors which satisfy a generalized version of Witten equation on the initial value surface. We show that the Witten-Nester charge, represented as an integral over the boundary of the initial value surface produces the Abbott-Deser-Tekin energy for asymptotically anti de Sitter spacetimes. We consider all values of the Chern-Simons coupling constant, including the critical value known as the chiral point, and study the cases of standard Brown-Henneaux boundary conditions, as well as their weaker version that allow a slower fall-off. Studying the Witten-Nester energy as a bulk integral over the initial value surface instead, we find a bound on the energy, and through it the sufficient condition for the positivity of the energy. In particular, we find that spacetimes of Petrov type $N$ that admit globally well defined solutions of the generalized Witten equation have positive energy.

\vspace{15pt}

\thispagestyle{empty}

\pagebreak


%

\tableofcontents


\newpage


\section{Introduction}


Topologically massive gravity (TMG) \cite{Deser:1982vy} with a
cosmological constant (CTMG)\cite{Deser:1982sv} is described by the action
\be
I= \frac1{16\pi G} \int d^3 x \sqrt{-g} \left( R-2\Lambda
 + \frac{1}{2\mu} \varepsilon^{\lambda\mu\nu}
\C^\rho_{\lambda\sigma} \left[ \partial_\mu\C^\sigma_{\rho\nu}
+\frac23 \C^\sigma_{\mu\tau} \C^\tau_{\nu\rho}\right] \right)\ ,
\label{1.1}
\ee
where the cosmological constant $\Lambda$ is negative.  This model provides an  attractive setting for studying several non-perturbative aspects of quantum gravity that are far more difficult to address in four and higher dimensions. The cosmological constant facilitates a black hole solution\cite{Banados:1992wn}, and the presence of the gravitational Chern-Simons (CS) term with critical strength \cite{LSS} may help in resolving certain obstacles that have been encountered in attempt to solve the theory exactly without the CS term \cite{Witten:1988hc,Witten:2007kt,Maloney:2007ud}.
One of the attractive features of the CTMG is that by AdS/CFT correspondence, and assuming appropriate boundary conditions, it admits a dual boundary CFT description. Indeed, there exist two copies of the Virasoro group that act as the boundary-condition preserving group, with central charges \cite{Brown:1986nw,Kraus:2005zm}
\be
c_L= \frac{3\ell}{2G} \left(1-\frac{1}{\mu\ell}\right)\ ,\qquad
c_R= \frac{3\ell}{2G} \left(1+\frac{1}{\mu\ell}\right)\ ,
\label{cc}
\ee
where $\ell$ is the AdS radius related to the cosmological constant as $\Lambda=-\ell^{-2}$. Assuming that $G>0$, we must have
\be
|\mu\ell| \ge 1\ ,
\ee
so that both central charges are non-negative. Without loss of generality we can choose $\mu$ to be positive by the parity transformation. Thus, we shall be interested in the regime $\mu\ell \ge 1$. However, the theory harbors a negative energy helicity 2 state for the conventional sign of the gravitational coupling constant $G>0$ \cite{Deser:1982vy,LSS}. While taking $G<0$ avoids this problem, it turns out to imply negative BTZ black hole mass. It has been proposed in \cite{LSS} that the instability problem in the case of $G>0$ can be circumvented by setting $\mu\ell=1$ in which case the negative energy graviton mode becomes identical to the already existing left-moving massless boundary graviton of the cosmological Einstein gravity, thereby ceasing to be a propagating bulk degree of freedom. The left central charge vanishes at $\mu\ell=1$, and the resulting theory subject to the standard Brown-Henneaux boundary conditions is called chiral gravity \cite{LSS}. It was conjectured in \cite{LSS} that the theory is chiral, in the sense that the physical states form representations of a single Virasoro algebra, and that it has positive energy.

Subsequently, the chirality conjecture was proven \cite{Strominger:2008dp,Carlip:2008qh}, though the positivity conjecture remains open. In particular, a propagating finite negative energy linearized mode at the chiral point was found \cite{Grumiller1}, which however, requires a weaker version \cite{Grumiller:2008es,Henneaux:2009pw} of the standard Brown-Henneaux boundary conditions. Finite negative energy modes obeying standard Brown-Henneaux boundary conditions were also found in \cite{Carlip:2008jk,Carlip:2008eq}, which however uses Poincar\'e patch which covers only a part of AdS, thus requiring further analysis to be conclusive. Subsequently, a linearized solution in the global coordinates which obeys the standard Brown-Henneaux boundary conditions was constructed as a descendant of a non-primary field \cite{Giribet:2008bw}. There seems to be an evidence, however, that this solution develops a logarithmic singularity in the next order in the weak field approximation\footnote{We thank Andy Strominger and Massimo Porrati for private communication about this problem.}. Finally, the chiral theory has also been analyzed in the Hamiltonian approach in which the counting of constraints reveals a single propagating degree of freedom even at the chiral point \cite{Park:2008yy,Grumiller2,Carlip:2008qh}. As emphasized in \cite{Carlip:2008qh}, however, even though this approach is nonperturbative, it does not address the boundary behavior.

The ultimate status of the chiral theory will require a full study of the AdS/CFT correspondence, including the characterization of an appropriate CFT, and verification of the self-consistency by showing the decoupling of the unwanted negative norm states \cite{Grumiller:2008es}. Alternatively, if the opposite overall sign in the action is chosen, then a superselection sector must be found in which the BTZ black holes are excluded, as has been suggested in \cite{Carlip:2008jk}.

Turning to the chiral theory, in addition to the positive energy BTZ black hole and the boundary graviton, it also admits the exact pp-wave solutions. These are locally AdS and apparently have zero energy. However, global considerations and how they may affect the energy require further study. Whether there exist other solutions of the chiral theory obeying the standard Brown-Henneaux boundary conditions is not known. Therefore, it clearly desirable to study the energy at the nonperturbative level, in search of a positive energy theorem. This is the main motivation for this work. In studying this problem, we shall employ a powerful tool introduced by Witten long ago \cite{Witten:1981mf}. This approach is inspired by the fact that global algebra of spinorial charges $Q$ in supergravity leads to the Hamiltonian $H={\rm Tr}\, Q^2$, and it was successfully used to provide a simple proof \cite{Witten:1981mf} of the  positive energy theorem in $4D$ for asymptotically Minkowskian spacetimes. The spinorial method for studying the energy in gravity was developed further in \cite{Nester:1982tr}.

The issue of positivity of the energy, or lack of thereof, was studied in $4D$ gravity theories with a class of curvature square terms in \cite{Boulware1,Boulware2} where it was shown that the $H={\rm Tr}\, Q^2$ argument is not sufficient by itself to guarantee positive energy, as possible presence of ghosts in higher derivative theories may give rise to a Hilbert space with an indefinite signature. While the CTMG has no ghost for $G<0$, and it has been proposed in \cite{Deser:1982sv} that the energy is positive in CTMG, based on the $H={\rm Tr}\, Q^2$ argument, the fact that the BTZ black hole has a negative energy for $G<0$ is a formidable problem that faces this proposal.

An attempt at calculating the Witten-Nester energy in CTMG has been made in \cite{gps}, where it was found not to have a definite sign in general. However, the gravitational CS term was essentially treated as a source term. Accordingly, the definition of the energy assumed in \cite{gps} differs from the existing ones based on Abbott-Deser approach \cite{Abbott:1981ff} applied to CTMG in \cite{Deser:2003vh} for asymptotically AdS spacetimes\footnote{We thank Andy Strominger for pointing out the need to modify the Noether supercurrent to remedy this problem.}
, and generalized in \cite{Bouchareb:2007yx} to accommodates spacetimes that are not asymptotically AdS.

Here we shall approach the energy problem by first constructing a Witten-Nester energy formula for CTMG that produces the Abbott-Deser-Tekin (ADT) energy when viewed as an integral over the boundary $\partial \Sigma$ of a spatial slice $\Sigma$. To achieve this, we develop a first order formulation of the locally supersymmetric extension of CTMG, since it provides the most convenient framework for constructing the supercurrent and superpotential associated with local supersymmetry. Exploiting the supersymmetric variation of the superpotential, we obtain an expression for conserved quantities, which in particular contains the energy which we refer to as the Witten-Nester energy, as an integral over the boundary $\partial\Sigma$ of the spatial slice. Converting this to a bulk integral over $\Sigma$, we obtain a bound on the Witten-Nester energy, and consequently a sufficient condition for its being positive.

In the calculations just outlined, the boundary conditions on the gravitational field play an important role in many respects, including the finiteness of the ADT charges. The standard Brown-Henneaux boundary conditions \cite{Brown:1986nw} were relaxed in \cite{Grumiller:2008es} and the most general boundary conditions invariant under the AdS group, asymptoting to AdS metric at infinity and yielding finite charges have been determined recently in \cite{Henneaux:2009pw}. Here, we examine the ADT charges for these most general boundary conditions, and show that indeed they are finite in all cases. These charges are closely related to the asymptotic symmetry charges discussed in \cite{Strominger:2008dp} and most recently in \cite{Henneaux:2009pw}. In agreement with \cite{Henneaux:2009pw}, we find that the charges associated with both null Killing vectors are nonvanishing even at the chiral point $\mu\ell=1$ (see Section 5).

In studying the Witten-Nester energy, we have found that it is convenient to generalize the well known Witten condition for a spinor which is a Dirac-like equation on the spacelike slice. We will describe this generalization for different values of the parameter $\mu$ in Section 3.3. At the chiral point $\mu\ell=1$ and with the standard Brown-Henneaux boundary conditions assumed, the generalized Witten equation takes the form
\be
\gamma^i e_i{}^\mu \left( {\nabla}_\mu -\frac{1}{2\ell}
\c_\mu -\frac{\ell^2}{4} C_{\mu\nu} \c^\nu \right) \eps = 0\ ,
\label{ws4}
\ee
where $C_{\mu\nu}$ is the  Cotton tensor. In Section 5.1 we show that the Witten-Nester energy is related to the ADT energy as  $E_{WN} = E_{ADT}$. Using \eq{ws4} and a Witten-Nester identity for the conserved charges as integral over the spacelike slice, we then find the following bound on the Witten-Nester energy
\be
 E_{WN} \ge - \frac{\ell^4}{32\pi G} \int_\Sigma \left(
C_\mu{}^\lambda C_{\lambda\nu}
 -\frac12 g_{\mu\nu} C^2 \right) u^\mu \,v^\nu\ da\ ,
\label{mpc4}
\ee
where $C^2 \equiv C_{\mu\nu} C^{\mu\nu}$, and $u^\mu$ is a unit timelike
vector perpendicular to the spatial slice $\Sigma$, and $v^\mu$  is
constructed out of the generalized Witten spinors as $v^\nu=
\bar{\epsilon}^1 \gamma^\nu \epsilon^1
+ \bar{\epsilon}^2 \gamma^\nu \epsilon^2$, which approaches
a timelike Killing vector on $\partial\Sigma$.
This result shows that to establish a
positive energy theorem by means of Witten-Nester type identity, we need
to know about the  Cotton tensor for spacetimes obeying the standard
Brown-Henneaux boundary conditions in the chiral theory. At present, the
only known exact solutions of this theory obeying the standard
Brown-Henneaux boundary conditions are spacetimes for which the Cotton
tensor vanishes. For these solutions, the bound is consistent with the
direct calculation of the ADT charges which shows that the energy
is positive. The utility of the result \eq{mpc4} lies in the fact that a
larger class of solutions may exist in which the Cotton tensor does not
vanish but has a positivity property for the simple algebraic integrand
we have found. A cursory look at the Petrov-like classification of $3D$
spacetimes shows that the integrand in \eq{mpc4} vanishes for Type $N$
metrics for which $C_{\mu\nu}$ is proportional to a product of two null
vectors. For all such solutions of the theory the energy is positive
provided that they obey the standard Brown-Henneaux boundary conditions,
and the generalized Witten equation \eq{ws4} has a globally well defined
solution. A positive energy theorem would result if one can show that
the metrics in the $\mu\ell=1$ theory that obey the standard
Brown-Henneaux boundary conditions are necessarily of Type $N$.
Whether this is the case remains to be seen. We shall come back to
this point in Section 6 where we will also discuss briefly other known
solutions without the  standard Brown-Henneaux asymptotics.

The outline of this paper is as follows. In Section 2 we shall describe the first order formulation of simple topologically massive supergravity. The use of supergravity is to facilitate the derivation of a Witten-Nester type expression for the energy in ordinary CTMG, and the first order formalism is developed because it provides the most convenient framework to achieve this\footnote{A first order formulation of topologically massive supergravity models has been proposed in \cite{Giacomini:2006dr} but they differ from the model we present here, as they contain a different set of fields.} .
In Section 3, we construct the Noether supercurrent and superpotential,
by following the procedure of \cite{silva} which builds on the work of
\cite{Regge:1974zd} and  which has been applied to ordinary supergravity
in \cite{Henneaux:1999ct}. This procedure is explained in Appendix
\ref{app:silva}. We then study the  integral of the supercurrent over a
spatial slice $\Sigma$, and by using Stokes' theorem, the integral over
its boundary $\partial\Sigma$. We compute the variation of the Noether
supercurrent and superpotential and propose a generalized version of the
Witten spinor equation, and derive a bound on the Witten-Nester
energy. In Section 4, we study the solution of the generalized Witten
spinor equation for the most general boundary conditions recently
proposed in \cite{Henneaux:2009pw}. In Section 5 we show that the
boundary integral representation of the Witten-Nester charges agree with
the ADT charges, and compute these charges. We also
compute the ADT charges directly for the general solutions of CTMG with vanishing Cotton tensor, and a class of solutions know as the chiral
pp-waves. Demanding consistency with our energy bound formula, we show that a globally well defined solution of the generalized Witten spinor equation must exist. We comment further on our results in Section 6, where we discuss further aspects of global issues in the definition of the Witten-Nester charges, implications of our energy bound for Petrov classes of $3D$ spacetimes, and the prospects of positive energy for $\mu\ell  >1$ in a nonperturbative setting.

\section{The Topologically Massive Supergravity}

\subsection{The Second-Order Formulation}


Simple topological massive supergravity was constructed by Deser and Kay \cite{DeserKay} and generalized to include a cosmological term by Deser \cite{Deser:1982sv}. This is $N=(1,0)$ supergravity since the supersymmetry parameter is a single Majorana spinor in the $(2,1)$ representation of the AdS group $SO(2,2)= SO(2,1)_L \times SO(2,1)_R$. The total Lagrangian, in our conventions (see Appendix A), is given by
\bea
e^{-1} \cL &=& R  +2 m^2 - 2 \varepsilon^{\mu\nu\rho} {\bar\psi}_\mu
D_\nu (\omega) \psi_\rho - m {\bar\psi}_\mu \c^{\mu\nu} \psi_\nu
\nn\w2
&& -\ft14
   \mu^{-1}\,\varepsilon^{\mu\nu\rho}\left( R_{\mu\nu}{}^{ab} \omega_{\rho
   ab}+ \ft23 \omega_\mu^{ab} \omega_{\nu b}{}^c\omega_{\rho ca} \right)
   -\mu^{-1} {\bar R}^\mu \c_\nu\c_\mu R^\nu\ , \label{1}
\eea
where $m = \ell^{-1}$ is the inverse radius of AdS.
We have set an overall factor of gravitational coupling constant
$16\pi G=1$, and used the following curvatures
\bea
R_{\mu\nu}{}^{ab} &=& \partial_\mu
\omega_\nu^{ab}+ \omega_\mu^{ac} \omega_{\nu c}{}^b- (\mu
\leftrightarrow \nu)\ ,\w2
R^\mu &=& \varepsilon^{\mu\nu\rho}
 D_\nu(\omega) \psi_\rho\ .
\eea
The covariant derivative of the gravitino in \eq{1} is defined as $D_\mu(\omega)\psi_\nu = \partial_\mu \psi_\nu +\ft14 \omega_\mu{}^{ab} \gamma_{ab} \psi_\nu$. The spin connection is not an independent field, but rather it is given by
\be
\omega_{\mu ab} = \omega_{\mu ab}(e) + \ft12\left( {\bar\psi}_\mu\c_a\psi_b -{\bar\psi}_\mu\c_b\psi_a +
{\bar\psi}_a\c_\mu\psi_b\right)\ ,
\label{ct}
\ee
and  $\omega_{\mu ab}(e)$ is the standard spin connection that solves the vanishing torsion equation $de^a+\omega^a{}_b e^b=0$. The action is invariant under the local supersymmetry transformations
\bea
\delta e_\mu^a &=& {\bar\epsilon}\c^a\psi_\mu\,,\nn\w2
\delta \psi_\mu &=& D_\mu(\omega)\epsilon -
   \ft12 m \c_\mu \epsilon\ . \label{susy}
\eea
Note that the $\mu$ dependent part of the action is separately invariant
under \eq{susy}. The field equations following from the Lagrangian
\eq{1} are \cite{gps}
\bea
&& \cG_{\mu\nu} + \mu^{-1} \, C_{\mu\nu} = 0\ ,\label{fe1}\w2
&& R^\mu +\ft12 m \c^{\mu\nu}\psi_\nu + \ft12 \mu^{-1}\, C^\mu =0\ ,\label{fe2}
\eea
up to fermionic terms in Einstein's equation, and cubic fermionic terms
in the gravitino field equation, and we have used the definitions
\bea
 \cG_{\mu\nu} &=& R_{\mu\nu}-\ft12 g_{\mu\nu} R - m^2 g_{\mu\nu}\ ,
 \label{cGdef}\w2
C_{\mu\nu} &=& \varepsilon_\mu{}^{\rho\sigma}
\nabla_\rho (R_{\sigma\nu} -\ft14 g_{\sigma\nu} R)\ , \label{Cotdef}\w2
C^\mu & = & \c^\rho \c^{\mu\nu} \nabla_\nu R_\rho -\varepsilon^{\mu\nu\rho}
\left(R_{\rho\sigma} -\ft14 g_{\rho\sigma} R \right) \c^\sigma \psi_\nu \ ,
\label{CCdef}
\eea
where $C_{\mu\nu}$, which is symmetric, traceless and divergence-free, is the Cotton tensor, and the vector-spinor $C^\mu$ is supersymmetric partner, the ``Cottino vector-spinor'' \cite{gps}. It follows from the bosonic field equation that
\bea
&& R = -6m^2\ ,
\label{f1}\w2
&& R_{\mu\nu} =  -2m^2 g_{\mu\nu} -\frac1{\mu} C_{\mu\nu}\ .
\label{f2}
\eea
%
%
%
So far we have described the topologically massive $N=(1,0)$ supergravity. The corresponding results for the $N=(0,1)$ theory are obtained by replacing $m \rightarrow -m$ everywhere.


\subsection{The First Order Formulation}


A first order formulation of the bosonic sector has been developed in \cite{Deser:1991qk,Carlip:1991zm,Carlip:2008qh}. Here we shall present a supersymmetric extension of this construction. The independent bosonic fields are $(e_\mu^a, f_\mu^a, A_\mu^a)$ and the independent fermionic fields are $(\psi_\mu,\eta_\mu)$. The vector spinor $\eta_\mu$ is Majorana.  The full Lagrangian in the first order formalism, the detailed derivation and properties of which will be discussed elsewhere, turns out to be
\bea
e^{-1} {\cal L} &=& {1\over \mu} \varepsilon^{\mu\nu\rho}
\left(\partial_\mu A_\nu^a +\ft13 \epsilon^a{}_{bc}
A_\mu^b A_\nu^c\right) A_{\rho a}
+ {1 \over \mu} {\bar\eta}^\mu \c_\nu\c_\mu \eta^\nu
\nn\w2
&&
- \frac{2}{\mu} \bar{\eta}^\mu \gamma_\nu \gamma_\mu R^\nu(A)
-4 {\bar\psi}_\mu R^\mu(A) +2(m^2-\mu^2)
-(m-\mu) {\bar\psi}_\mu\c^{\mu\nu}\psi_\nu
\nn\w2
&& + \varepsilon^{\mu\nu\rho} \left( D_\mu(A) e_{\nu a}
+\mu \epsilon_{abc} e_\mu^b e_\nu^c
-\ft12{\bar\psi}_\mu \c_a \psi_\nu\right) f_\rho^a\ ,
\eea
where
\bea
R^\mu (A) &=& \varepsilon^{\mu\nu\rho} ( \partial_\nu +\ft12 A_\nu^a\c_a )\,\psi_\rho\ ,
\nn\w2
D_\mu (A) e_{\nu a} &=& \partial_\mu e_{\nu a} + \epsilon_{abc} A_\mu^b e_\nu^c\ .
\eea
An important advantage of this formulation is that all the terms in the Lagrangian contain at most a single derivative.

The equations of motion for $(f_\mu{}^a, A_\mu^a, e_\mu^a)$, respectively, and up to fermionic bilinears, are
\bea
&&  D_{[\mu} (A) e_{\nu] a} +\mu \epsilon_{abc} e_\mu^b e_\nu^c  =0 \ ,
\label{eom1}\w2
&& F_{\mu\nu a} + \mu \epsilon_{abc} f_{[\mu}{}^b e_{\nu]}{}^c=0\ ,
\label{eom2}\w2
&& D_{[\mu}(A) f_{\nu]a} + 2\mu \epsilon_{abc} f_{[\mu}^b e_{\nu]}^c
-(m^2-\mu^2) \varepsilon_{\mu\nu\rho} e^\rho_a=0\ ,
\label{eom3}
\eea
where
\be
F_{\mu\nu}{}^a = \partial_\mu A_\nu^a -\partial_\nu A_\mu^a
+ \epsilon^{abc} A_{\mu b} A_{\nu c}\ .
\ee
The first equation is solved by
\be
A_\mu^a = \omega_\mu^a (e) -\mu e_\mu^a\ ,
\label{A}
\ee
where $\omega_\mu^a = \ft12 \epsilon_{abc} \omega_\mu{}^{bc}$. Substituting this result into \eq{eom2}, it is straightforward to solve for $f_\mu{}^a$, and one finds
\be
f_\mu{}^a = -\ft2{\mu} \left[ R_\mu{}^a -\ft14 e_\mu^a R
+ \ft12 \mu^2 e_\mu^a \right] \ ,
\label{f}
\ee
where $R_\mu{}^a = R_{\mu\nu} e^{\nu a}$ is the Ricci tensor. Substituting
this result into \eq{eom3} then gives precisely the equation of motion
(\ref{fe1})
that was found in the second order formalism.

Turning to the fermionic fields, up to cubic fermionic terms the field
equations for $\eta_\mu$ and $\psi_\mu$, respectively are
\bea
&& \gamma^\nu\gamma^\mu \left(\eta_\nu - R_\nu(A)\right)=0\ ,
\label{fequation1}\w2
&& R^\mu(A)
+ \frac{1}{4} (m-\mu) \gamma^{\mu\nu} \psi_\nu
+ \frac{1}{8} \varepsilon^{\mu\nu\rho} \gamma_a \psi_\nu f_\rho^a
 - \frac{1}{4\mu} \varepsilon^{\mu\nu\rho} D_\rho(A)
\left( \gamma_\sigma \gamma_\nu \eta^\sigma\right)=0\ .
\label{fequation2}
\eea
The first equation is readily solved to give
\be
\eta^\mu = R^\mu (A)\ .
\label{eta}
\ee
Using this relation, as well as \eq{A} and \eq{f} in the second
equation, after some algebra we find  the gravitino field equation
(\ref{fe2}),
thereby completing the proof that the field equations in the first
order formalism are classically equivalent to those obtained in the
second order formalism.

The supersymmetry transformations for the dreibein and the gravitino are
\bea
\delta e_\mu^a &=& {\bar\epsilon} \c^a \psi_\mu\ ,
\nn\w2
\delta \psi_\mu
&=&  D_\mu(A)\epsilon - \ft12 (m-\mu) \c_\mu \epsilon\ .
\label{deltas0}
\eea
The supersymmetry transformation of the remaining fields $(A_\mu^a, f_\mu^a, \eta^\mu)$ can be obtained by using the expressions \eq{A}, \eq{f} and \eq{eta}. They will be fully provided in \cite{st} but for our purposes here we need to know the terms that involve the derivative of the supersymmetry parameter. We find that
\bea
\delta \eta^\mu & = & - \frac{1}{2} (m-\mu) \varepsilon^{\mu\nu\rho}
\gamma_\rho \partial_\nu \epsilon + Z^\mu \epsilon\ ,
\nn\w2
\delta f_\mu^a & = & - \frac{1}{\mu} \partial_\mu \bar{\epsilon}
\left( \gamma^b \gamma^a R_b(\omega) - m \psi^a \right) + \bar\epsilon \chi_\mu^a\ ,
\nn\w2
\delta A_\mu^a &=& \bar\epsilon \lambda_\mu^a\ ,
\label{deltas}
\eea
where $Z^\mu, \chi_\mu^a, \lambda_\mu^a$ are certain algebraic expressions \cite{st} depending on the fields of the theory whose precise form we do not need for our purposes here.


\section{The Witten-Nester Energy}


\subsection{The Noether Supercurrent and Superpotential}


The Witten-Nester charges, one of which is the energy, are defined as
\be
Q_{WN} = \int_\Sigma \delta_{\eps_2} J^\mu_{\eps_1} d\Sigma_\mu
= \int_\Sigma \nabla_\nu \left(\delta_{\eps_2} U^{\mu\nu}_{\eps_1}
\right) d\Sigma_\mu \ ,
\label{vi}
\ee
where $\Sigma$ is a spacelike initial value surface, $J^\mu_{\eps}$ is
the Noether current associated with local supersymmetry transformations
with parameter $\eps$, and $U_\epsilon^{\mu\nu}$ is the superpotential
$J_\epsilon^\mu = \nabla_\nu U_\epsilon^{\mu\nu}$.
Using Stokes' theorem, and assuming that the surface $\Sigma$ has a
{\it single\ boundary}, and assuming that the bulk integral
$\int_\Sigma \delta_{\eps_2} J^\mu_{\eps_1} d\Sigma_\mu$ is
{\it finite},  $Q_{WN}$ can be re-expressed as
\be
Q_{WN}= \int_{\partial\Sigma} \left(\delta_{\eps_2}
U^{\mu\nu}_{\eps_1}\right) d\Sigma_{\mu\nu}\ .
\label{si}
\ee
We expect a relation between this expression and the ADT charges given in Appendix \ref{app:adt}. If such a relation exists, it will then enable us to study its positivity property by exploiting the expression \eq{p2}. This is the strategy pioneered by Witten \cite{Witten:1981mf} in his proof of the positive energy theorem in $4D$ gravity in asymptotically Minkowskian spacetimes.

In this section we shall determine the Noether supercurrent $J^\mu_\eps$
and superpotential $U_\eps^{\mu\nu}$. We shall also determine the
variation of the supercurrent under local supersymmetry, and examine
the integral \eq{vi} over $\Sigma$. In Section 4 we shall examine the
integral \eq{si} over $\partial\Sigma$.

Using the formula \eq{silva} explained in Appendix \ref{app:silva} letting $\xi^a \rightarrow \eps$ and reading off $\Delta^\mu_a$ from \eq{deltas0}, \eq{deltas}, we construct the Noether supercurrent associated with the local supersymmetry of the topologically massive $N=(1,0)$ supergravity as
\bea
J^{\mu L}_\eps  &=& \nabla_\nu U_\epsilon^{\mu\nu}
+ \bar{\epsilon} \frac{\delta {\cal L}}{\delta \bar{\psi}_\mu}
+ \frac{1}{2} (m-\mu) \varepsilon^{\nu\mu\rho} \bar{\epsilon}
\gamma_\rho \frac{\delta {\cal L}}{\delta \bar{\eta}^\nu}
- \frac{1}{\mu} \bar{\epsilon} \left( \gamma^b \gamma^a R_b(\omega)
- m \psi^a \right)
\frac{\delta {\cal L}}{\delta f_\mu^a}
\nn\w2
&=& \nabla_\nu U_\epsilon^{\mu\nu}
- 8 \bar{\epsilon} \left( R^\mu(A)
+ \frac{1}{4} (m-\mu) \gamma^{\mu\nu} \psi_\nu \right.
+ \frac{1}{8} \varepsilon^{\mu\nu\rho} \gamma_a \psi_\nu f_\rho^a
\nn\w2
&&
\left. - \frac{1}{4\mu} \varepsilon^{\mu\nu\rho} D_\rho(A)
\left( \gamma_\sigma \gamma_\nu \eta^\sigma \right)
\right)
+ \frac{m-\mu}{\mu} \varepsilon^{\nu\mu\rho} \bar{\epsilon}
\gamma_\rho \gamma_\sigma \gamma_\nu \left( \eta^\sigma
- R^\sigma(A) \right)
\nn\w2
&& - \frac{1}{\mu} \bar{\epsilon} \left(
\gamma^b \gamma^a R_b(\omega) - m \bar{\epsilon} \psi^a \right)
\varepsilon^{\rho\sigma\mu} \left(
D_\rho(A) e_{\sigma a} + \mu \epsilon_{acd} e_\rho^c e_\sigma^d
- \frac{1}{2} \bar{\psi}_\rho \gamma_a \psi_\sigma \right).
\eea
Its variation is
\be
\delta J^{\mu L}_\eps = \nabla_\nu \left[ \delta U_\epsilon^{\mu\nu}
- 8 \varepsilon^{\mu\nu\rho} \bar{\epsilon} \left( \delta\psi_\rho
+ \frac{1}{4\mu} \gamma_\sigma \gamma_\rho \delta\eta^\sigma \right)
- \frac{2(m-\mu)}{\mu} \varepsilon^{\mu\nu\rho}
\bar{\epsilon} \delta\psi_\rho
\right] + \cdots,
\ee
where $\cdots$ denotes terms which do not contain $\partial\delta\phi$,
where $\phi$ is a generic field. By requiring that $\delta J^\mu$ does
not have $\partial\delta\phi$ terms, as explained in Appendix \ref{app:silva},
we obtain
\be
U^{\mu\nu L}_\epsilon
= 8 \varepsilon^{\mu\nu\rho} \bar{\epsilon} \left(
\psi_\rho
+ \frac{1}{4\mu} \gamma_\sigma \gamma_\rho \eta^\sigma \right)
+ \frac{2(m-\mu)}{\mu} \varepsilon^{\mu\nu\rho} \bar{\epsilon}
\psi_\rho\ .
\ee
Using the equation of motion for $\eta$, this becomes
\be
U^{\mu\nu L}_\epsilon = 4 \left( 1 + \frac{m}{2\mu} \right)
\varepsilon^{\mu\nu\rho} \bar{\epsilon} \psi_\rho
+ \frac{2}{\mu} \varepsilon^{\mu\nu\rho} \bar{\epsilon}
\gamma_\sigma \gamma_\rho R^\sigma(\omega)\ .
\label{u}
\ee
Thus, the Noether current is given by
\be
J^{\mu L}_\eps = \nabla_\nu U^{\mu\nu L}_\epsilon\ .
\label{nc}
\ee
The Witten-Nester charges
\be
Q_{WN}^L  = \int_\Sigma \delta_{\eps_2} J^{\mu L}_{\eps_1} d\Sigma_\mu
= \int_{\partial\Sigma} \left(\delta_{\eps_2}
U^{\mu\nu L}_{\eps_1}\right) d\Sigma_{\mu\nu}
\label{vsi}
\ee
generate the $SO(2,1)_L$ subalgebra of the $AdS_3$ algebra
$SO(2,1)_L \times SO(2,1)_R$. The $SO(2,1)_R$ charges are obtained
by making the replacement $m \rightarrow -m$ in \eq{u} which yields
$U^{\mu\nu R}_\eps$. These expressions hold for any value of $\mu$
including  the chiral point $\mu=m$. Note also that we have set $16\pi G=1$
in \eq{vsi}.


\subsection{Witten-Nester Formula and the Gravitino Field Equation}


Before examining the Witten-Nester charges \eq{vsi} let us make a
comment on a relation between the supercurrent \eq{nc} and the gravitino
field equation \eq{fe2} following the approach in \cite{Abbott:1981ff}.
Here, we consider a metric which admits a Killing spinor satisfying
\be
\left( \nabla_\mu - \ft12 m \gamma_\mu \right) \epsilon = 0\ .
\label{kse}
\ee
The integrability condition of this equation is
\be
C_{\mu\nu} \gamma^\nu \epsilon = 0
\label{intcond}
\ee
We take the linearized gravitino field equation \eq{fe2} and multiply
it by a Killing spinor $\bar{\epsilon}$.
We find for the first two terms in \eq{fe2}
\be
\bar{\epsilon} \left( R^\mu + \ft12 m \gamma^{\mu\nu} \psi_\nu \right)
= \nabla_\nu \left( \varepsilon^{\mu\nu\rho }\bar{\epsilon} \psi_\rho
\right)\ ,
\ee
and for the third term
\be
\bar{\epsilon} C^\mu
= \nabla_\nu \left( \bar{\epsilon} \gamma^\rho \gamma^{\mu\nu} R_\rho
+ m \varepsilon^{\mu\nu\rho} \bar{\epsilon} \psi_\rho \right)
- \varepsilon^{\mu\nu\rho} \left( R_{\rho\sigma}
- \ft14 g_{\rho\sigma} R + \ft12 m^2 g_{\rho\sigma} \right)
\bar{\epsilon}\gamma^\sigma \psi_\nu\ .
\label{efe2}
\ee
Using the gravitational field equation \eq{fe1} the second term in
\eq{efe2} is proportional to \eq{intcond}, which vanishes.
Thus we find that the gravitino field equation multiplied by a Killing
spinor gives the supercurrent we obtained by the Noether method
\be
\bar{\epsilon} \left( R^\mu + \ft12 m \gamma^{\mu\nu} \psi_\nu
+ \ft12 \mu^{-1} C^\mu \right)
= \ft14 \nabla_\nu U_\epsilon^{\mu\nu L}\ ,
\label{fe2u}
\ee
where $U^{\mu\nu L}_\epsilon$ is the superpotential in \eq{u}.
However, note that to derive \eq{fe2u} we assumed that the metric admits
a Killing spinor and used the Killing spinor $\epsilon$ in the superpotential
$U_\epsilon^{\mu\nu L}$. In the Noether method above, the supercurrent \eq{nc} was obtained for the general metric and spinor $\epsilon$.


\subsection{The Boundary Integral}


To study the boundary integral representation of the Witten-Nester
charge (\ref{si})
we compute the supersymmetry variation of $U^{\mu\nu}_\eps$. Computing
the $N=(1,0)$ supersymmetry variation of $U^{\mu\nu L}_\eps$ defined
in \eq{u}, we find
\bea
Q_{WN}^L &=&
 \int \varepsilon^{\mu\nu\lambda} \left[
4\left(1+\frac{m}{\mu}\right) {\bar\eps}_2 \hat{\nabla}_\lambda^L \eps_1
+ \frac{2}{\mu} {\bar\eps}_2 \c^\sigma\eps_1\left(\cG_{\lambda\sigma}
- \frac12 g_{\lambda\sigma} \cG\right)\right] d\Sigma_{\mu\nu}\ ,
\label{ab}
\eea
where we have set $16\pi G=1$, the commuting spinors $\eps_1$ and
$\eps_2$ take their asymptotic values which can be taken to be any one
of the Killing spinors of AdS $\eps_K^{A}$ discussed in Appendix
\ref{app:killing}, and
\be
\hat{\nabla}^L_\mu \epsilon = \left( \nabla_\mu
- \frac{1}{2} m \gamma_\mu \right) \epsilon\ .
\ee
Depending on which Killing spinors are utilized in \eq{ab}, the Killing vectors defined as in \eq{k1} emerge, and the charges $Q_{WN}^L$ correspond to these Killing vectors. To keep the notation simple, we do not exhibit the corresponding labels of the Witten-Nester charges.

Similarly, the $N=(0,1)$ supersymmetry variations of $U^{\mu\nu R}_\eps$ leads to $Q_{WN}^R$ which can be obtained from $Q_{WN}^L$ by letting $m \rightarrow -m$, and the spinors $\eps_1$ and $\eps_2$ take their asymptotic values which can be taken to be any one of the Killing spinors $\eps_K^{\dot A}$ discussed in Appendix \ref{app:killing}. This time, the the Killing vectors defined as \eq{k2} emerge, and the resulting charges $Q_{WN}^R$ correspond to these Killing vectors.

The relation between the Witten-Nester charges $Q_{WN}^{L,R}$ described
above and the ADT charges defined in Appendix \ref{app:adt} will be
established in Section 5.1.


\subsection{The Bulk Integral and a Bound on the Witten-Nester Energy}


To study the bulk integral representation of the Witten-Nester charges
(\ref{vi}) we evaluate the $N=(1,0)$ supersymmetry variation of the
Noether current defined in \eq{nc}. After some calculations we obtain
\bea
\delta_{\eps_2} J^{\mu L}_{\eps_1}
&=& 4\left(1+\frac{m}{\mu} \right) \hat{\nabla}^L_\nu
\bar{\epsilon}_1 \gamma^{\mu\nu\rho}
\hat{\nabla}^L_\rho \epsilon_2
+ \bar{\epsilon}_1 \gamma^\rho \epsilon_2 \left[
2\left(1+\frac{m}{2\mu}\right) \cG_\rho{}^\mu
+\frac2{\mu} C_\rho{}^\mu \right]
\label{exp1}\w2
&& + \frac{2}{\mu} \varepsilon^{\mu\nu\rho} \left( R_{\rho\sigma}
- \frac{1}{4} g_{\rho\sigma} R
+ \frac{1}{2} m^2 g_{\rho\sigma} \right)
\left[
\hat{\nabla}^L_\nu \bar{\epsilon}_1 \gamma^\sigma \epsilon_2
+ \bar{\epsilon}_1 \gamma^\sigma \hat{\nabla}^L_\nu \epsilon_2
- \frac{1}{2} m \bar{\epsilon}_1 \gamma_\nu{}^\sigma \epsilon_2
\right]\ .
\nn
\eea
The result \eq{exp1} can be simplified by using the full Einstein
equation \eq{fe1} and its consequences \eq{f1}, \eq{f2} to give
\be
Q_{WN}^L
= \int_{\Sigma} d\Sigma_\mu\,\left[4\left(1+\frac{m}{\mu} \right)
\hat{\nabla}^L_\nu
\bar{\epsilon}_1 \gamma^{\mu\nu\rho} \hat{\nabla}^L_\rho \epsilon_2
- \frac{2}{\mu^2} \varepsilon^{\mu\nu\rho} C_{\rho\sigma}
\left( \hat{\nabla}^L_\nu \bar{\epsilon}_1 \gamma^\sigma \epsilon_2
+ \bar{\epsilon}_1 \gamma^\sigma \hat{\nabla}^L_\nu \epsilon_2 \right) \right]\ .
\label{master1}
\ee
This expression can be expressed in various alternative forms. One possible strategy is to extract a boundary integral by manipulating the last term by means of partial integration and use of equations  of motion. In this way, after some algebra we find
\be
Q_{WN}^L
= \int_{\Sigma} d\Sigma_\mu\, \left[  4\left(1+\frac{m}{\mu} \right)
\left(\hat{\nabla}^L_\nu
\bar{\epsilon}_1 \gamma^{\mu\nu\rho} \hat{\nabla}^L_\rho \epsilon_2
-\frac1{2\mu} C^{\mu\nu} \bar{\epsilon}_1 \gamma_\nu \epsilon_2 \right)
-\frac{2}{\mu^2} \nabla_\nu \left( \varepsilon^{\mu\nu\rho} C_{\rho\sigma}
\bar{\epsilon}_1 \gamma^\sigma \epsilon_2 \right)\right]\ .
\label{dj}
\ee
Alternatively, recalling that $\mu \ge m$, it is convenient to express
\eq{master1} as
\be
Q_{WN}^L
= \int_{\Sigma} d\Sigma_\mu\, \left[ 4\left(1+\frac{m}{\mu} \right)
\widetilde{\nabla}^L_\nu \bar{\epsilon}_1 \gamma^{\mu\nu\rho}
\widetilde{\nabla}^L_\rho \epsilon_2
+ \frac{1}{\mu^3(\mu+m)} \varepsilon^{\mu\nu\rho}
C_{\nu\sigma} C_{\rho\lambda} \varepsilon^{\sigma\lambda\alpha}
\bar{\epsilon}_1 \gamma_\alpha \epsilon_2 \right]\ ,
\label{p2}
\ee
where
\be
{\widetilde\nabla}^L_\mu \eps := \left( \nabla_\mu
-\frac12 m \c_\mu - \frac{1}{2\mu(\mu+m)} C_{\mu\nu}\c^\nu \right) \eps\ .
\label{ws33}
\ee
In this case, we impose a generalized version of the Witten spinor
condition as
\be
\gamma^i {\widetilde\nabla}^L_i \eps=0\ ,
\label{ws2}
\ee
where ${\widetilde\nabla}^L_i
= e_i{}^\mu {\widetilde\nabla}^L_\mu$ ($i=1,2$).
Next, we use the identity $\c^{0ij} = -\c^{i} \c^0 \c^{j}
- \delta^{ij} \c^0$ and the definition of the Dirac conjugate
\eq{dconj} to obtain the result
\bea
Q_{WN}^L &=& 4\left(1+\frac{m}{\mu}\right) \int_\Sigma
\left({\widetilde\nabla}^L_i \eps_1\right)^\dagger
\left({\widetilde\nabla}^L_i\eps_2\right)\, e_0{}^\mu d\Sigma_\mu
\nn\w2
&& - \frac{2}{\mu^3(\mu+m)} \int_\Sigma ( C^{\mu\lambda} C_{\lambda\nu}
-\frac12 \delta^\mu_\nu C^2 ) (\bar{\epsilon}_1 \gamma^\nu \epsilon_2)\,
d\Sigma_\mu \ ,
\label{mpc2}
\eea

\noindent where we have expressed the product of two $\eps$ symbols
in terms of Kronecker deltas and used the notation
$C^2 \equiv C^{\mu\nu} C_{\mu\nu}$. The first term is manifestly
positive when $\epsilon_1 = \epsilon_2$, and therefore we have the bound
\be
\mu \ge m: \qquad  Q_{WN}^L \ge - \frac{2}{\mu^3(\mu+m)}
\int_\Sigma da\, X_{\mu\nu}\, u^\mu v^{(+)\nu}\ ,
\label{mpc}\w2
\ee
where we have set $16\pi G=1$,
\be
X_{\mu\nu} := C_\mu{}^\lambda C_{\lambda\nu}
- \frac12 g_{\mu\nu} C^2\ ,
\label{xdef}
\ee
and $d\Sigma_\mu = u_\mu da$, with $u_\mu$ representing a timelike vector that specify the orientation of the area element $da$ on the initial spacelike surface, and we have the vector
\be
v^{(+)\nu} := \bar{\epsilon} \gamma^\nu \epsilon\ ,
\label{LSpin}
\ee
where the spinors are the solutions of \eq{ws2} asymptoting to  the
Killing spinors $\eps_K^A$ (see Appendix D).

Next we turn to $Q^R_{WN}$. The relevant equations for these charges are obtained from \eq{ws2}, \eq{ws33}, \eq{mpc2} and \eq{mpc} by simply letting $m \rightarrow -m$, and in \eq{LSpin} taking the spinors to approach the dotted Killing spinors asymptotically. As a result, we have the bound:
\be
\mu > m:  \qquad  Q_{WN}^R \ge - \frac{2}{\mu^3(\mu-m)} \int_\Sigma da\,
X_{\mu\nu}\, u^\mu v^{(-)\nu}\ ,
\label{mp4}
\ee
where the vector $v^{(-)\nu}$ is the bilinear \eq{LSpin} in which the
spinors now obey $\gamma^i {\widetilde\nabla}^R_i \eps=0$
and asymptote to the Killing spinors $\eps_K^{\dot A}$ (see Appendix D).
The chiral point $\mu=m$ clearly requires special care. In this case, it
is natural to directly evaluate the expression that results from \eq{dj}
by replacing $m \rightarrow -m$. Setting $\mu=m$, the first two terms
cancel and we find
\be
Q_{WN}^R
= -\frac{2}{\mu^2}\int_{\Sigma} d\Sigma_\mu\,
 \nabla_\nu \left( \varepsilon^{\mu\nu\rho} C_{\rho\sigma}
\bar{\epsilon}_1 \gamma^\sigma \epsilon_2 \right)\ .
\label{djr2}
\ee
Using the field equation $C_{\mu\nu}=- \mu \cG_{\mu\nu}$, and noting
\eq{aterm} and \eq{2ndterm}, we see upon comparison with the expression
\eq{E+J}, that the right hand side of \eq{djr2} being a boundary
integral, it does produce, as to be expected, the ADT expression \eq{E+J}.
Furthermore, as we shall show in Section 5.2, this particular ADT
charge vanishes if standard Brown-Henneaux boundary conditions
are imposed. Thus, we have
\be
\mu = m: \qquad Q_{WN}^R=0 \qquad  \mbox{(standard Brown-Henneaux b.c.)}\ .
\label{mp5}
\ee
%


\section{The Witten Spinor}


To study the Witten-Nester identity \eq{mpc2} we need to know the asymptotic solution of the generalized Witten spinor equation \eq{ws2} with appropriate boundary conditions. We shall be primarily interested in the standard Brown-Henneaux boundary conditions to be described below, and the chiral point $\mu\ell=1$. Nonetheless, it is instructive to see the consequences of imposing the weakest possible boundary conditions allowed for different values of $\mu$ as well. With this in mind, we shall study the asymptotic solutions of the generalized Witten equation for the most general boundary conditions compatible with the AdS asymptotics.


\subsection{The HMT Boundary Conditions}


The most general boundary conditions that are invariant under the AdS
group, and asymptoting to AdS metric at infinity and yielding finite
charges have been determined recently by Henneaux, Mart\'inez and
Troncoso (HMT) \cite{Henneaux:2009pw}.
We expand the gravitational field around the $AdS_3$
vacuum solution given by
\be
d\bar{s}^2 = \ell^2 \left[
- \cosh^2\rho \, d\tau^2 + \sinh^2\rho \, d\phi^2 + d\rho^2 \right]
\label{ads}
\ee
and define the deviation fields as
\bea
e_\mu{}^a &=& \bar{e}_\mu{}^a + \frac{1}{2} h_{\mu}{}^a\ ,
\nn\w2
g_{\mu\nu} &=& \bar{g}_{\mu\nu} + h_{\mu\nu} + {\cal O}(h^2)\ ,
\label{eg}
\eea
where we choose $h_{\mu\nu} \equiv \bar{e}_{\nu a} h_{\mu}{}^a$ to
be symmetric $h_{\mu\nu} = h_{\nu\mu}$ by local Lorentz transformation.
The coordinates have the ranges $ 0\le \rho <\infty$, $0 \le \phi <2\pi$,
and $ 0 \le \tau < 2\pi$ for $AdS$, and $-\infty < \tau < \infty$ for
its universal cover. Another useful form of the metric is obtained
by defining $t=\ell\tau$, $r=\ell\sinh\rho$ and takes the form
\be
d\bar{s}^2 = - \left( 1 + m^2r^2 \right) dt^2 + r^2 d\phi^2
+ \frac{dr^2}{1+m^2r^2}\ .
\label{global}
\ee

Depending on the value of the parameter $\mu\ell$, there exist different
possible boundary conditions on the components of
$h_{ab}={\bar e}_a{}^\mu {\bar e}_b{}^\nu h_{\mu\nu}$.
These boundary conditions are (see Appendix \ref{app:conv} for notation):
\bea
|\mu\ell| > 1: \qquad\qquad  h_{++} &=& e^{-2\rho} f_{++}  +\cdots
\nn\w2
h_{+-} &=& e^{-2\rho} f_{+-} + \cdots
\nn\w2
h_{22} &=& e^{-2\rho} f_{22} + \cdots
\nn\w2
h_{--} &=& e^{-2\rho} f_{--} +\cdots
\nn\w2
h_{+2} &=& e^{-3\rho} f_{+2} +\cdots
\nn\w2
h_{-2} &=& e^{-3\rho} f_{-2} + \cdots
\label{sbc}
\eea
where $f_{ab}$ depend only on $\tau$ and $\phi$.
These are the standard Brown-Henneaux boundary conditions
\cite{Brown:1986nw}. The newly established boundary conditions arise
when one considers the parameter range $|\mu\ell| <1$.
They are \cite{Henneaux:2009pw}
\bea
0< |\mu\ell| <1: \qquad\qquad  h_{++} &=& e^{-2\rho} f_{++}  +\cdots
\nn\w2
h_{+-} &=& e^{-2\rho} f_{+-} + \cdots
\nn\w2
h_{22} &=& e^{-2\rho} f_{22} + \cdots
\nn\w2
h_{--} &=& e^{-(1+\mu\ell)\rho} \tf_{--} + e^{-2\rho} f_{--} +\cdots
\nn\w2
h_{+2} &=& e^{-3\rho} f_{+2} +\cdots
\nn\w2
h_{-2} &=& e^{-(2+\mu\ell)\rho} \tf_{-2} + e^{-3\rho} f_{-2} + \cdots
\label{nbc}
\eea
where $f_{ab}$ and $\tilde{f}_{ab}$ depend only on $\tau$ and $\phi$. Note
that the $f$-terms are the standard deviations from $AdS$ given in
\eq{sbc} and the $\tf$-terms represent the generalizations of them with
slower fall-off. These boundary conditions have been called ``negative
chirality'' boundary conditions since only $h_{-2}$ and $h_{--}$ have a
slower fall-off terms.  For the same range $0< |\mu\ell| <1$, there are
also ``positive chirality'' boundary conditions which are obtained from
the ones in \eq{nbc} by replacing everywhere
the indices $+ \leftrightarrow -$.

The remaining possible boundary conditions described in
\cite{Henneaux:2009pw} arise at the so-called chiral point at which
$|\mu\ell|=1$ and had been already proposed in \cite{Grumiller:2008es},
and they take the form:
\bea
\mu\ell=1: \qquad\qquad  h_{++} &=& e^{-2\rho} f_{++}  +\cdots
\nn\w2
h_{+-} &=& e^{-2\rho} f_{+-} + \cdots
\nn\w2
h_{22} &=& e^{-2\rho} f_{22} + \cdots
\nn\w2
h_{--} &=& \rho \, e^{-2\rho} \tf_{--} + e^{-2\rho} f_{--} +\cdots
\nn\w2
h_{+2} &=& e^{-3\rho} f_{+2} +\cdots
\nn\w2
h_{-2} &=& \rho \, e^{-3\rho} \tf_{-2} + e^{-3\rho} f_{-2} + \cdots
\label{cbc}
\eea
Finally, for $\mu\ell=-1$, the allowed boundary conditions are obtained
from these by interchanging $+ \leftrightarrow -$. An observation which
will be useful later is that the boundary conditions for $\mu\ell=1$
(and similarly for $\mu\ell = -1$) can be obtained from those in
\eq{nbc} by first replacing
\bea
\tilde{f}_{--} & \rightarrow & \frac{\tilde{f}_{--}}{1-\mu\ell}, \qquad
f_{--} \ \rightarrow \ - \frac{\tilde{f}_{--}}{1-\mu\ell}
+ f_{--},
\nn\w2
\tilde{f}_{-2} & \rightarrow & \frac{\tilde{f}_{-2}}{1-\mu\ell}, \qquad
f_{-2} \ \rightarrow \ - \frac{\tilde{f}_{-2}}{1-\mu\ell}
+ f_{-2}\ ,
\label{rule}
\eea
and then taking the limit $\mu\ell \rightarrow 1$. Remarkably, all the
known exact solutions of CTMG with asymptotically AdS behavior, which
are essentially the chiral pp-waves and exist for both $|\mu\ell| >1$
as well as $|\mu\ell| =1$ \cite{AyonBeato:2004fq,Carlip:2008jk,gps},
have exactly the above boundary behavior.

While all the above boundary conditions are allowed in the sense
mentioned earlier, the requirement that the two copies of the Virasoro
algebra arising as the asymptotic symmetry algebra have {\it positive}
central charges imposes the condition $|\mu\ell| \ge 1$.

For later use let us show that the coefficient functions $f_{ab}$
appearing in the HMT boundary conditions satisfy
\be
f_a{}^a = 2 f_{+-} + f_{22} = 0\ .
\label{ftrace}
\ee
To achieve this we use the fact that the deviations $h_{ab}$
asymptotically satisfy the linearized field equation derived from \eq{fe1}
\be
\delta {\cal G}_{ab} + \frac{1}{\mu} \varepsilon_a{}^{cd}
\bar{\nabla}_c \delta {\cal G}_{bd} = 0\ ,
\label{linfe}
\ee
where
\be
\delta {\cal G}_{ab}
= \frac{1}{2} \left( - \bar{\nabla}^2 h_{ab}
- \bar{\nabla}_a \bar{\nabla}_b h
+ \bar{\nabla}^c \bar{\nabla}_a h_{bc}
+ \bar{\nabla}^c \bar{\nabla}_b h_{ac} \right)
+ 2m^2 h_{ab}\ .
\label{lincg}
\ee
By taking trace of \eq{linfe} we obtain
\be
- \bar{\nabla}^2 h + \bar{\nabla}^a \bar{\nabla}^b h_{ab} + 2m^2 h = 0\ ,
\label{trlin}
\ee
where $h = h_a{}^a$.
Substituting the HMT boundary conditions into \eq{trlin} the first two
terms are shown to be of higher order than ${\cal O}(e^{-2\rho})$, while
the last term is proportional to $h = e^{-2\rho} f_a{}^a + \cdots$.
Therefore, we obtain \eq{ftrace}.


\subsection{The Asymptotic Solution of the Generalized Witten Equation}


With the boundary conditions at hand, we are ready to examine the
generalized Witten equation \eq{ws2}.\footnote{In the case of ordinary
AdS gravity without the gravitational CS term in $3D$, the Witten
equation has been studied in arbitrary dimensions in \cite{adiherif}.}
In the $AdS_3$ background (\ref{ads}), the only nonvanishing components
of the spin connection are given by
$\bar{\omega}_\tau{}^{02} = \sinh\rho$,
$\bar{\omega}_\phi{}^{12} = \cosh\rho$.
Equivalently, the only nonvanishing components of
$\bar{\omega}_c{}^{ab} = \bar{e}_c{}^\mu \bar{\omega}_\mu{}^{ab}$
are
\be
\bar{\omega}_+{}^{+2} = \bar{\omega}_-{}^{-2} = m \coth2\rho\ .
\label{om2}
\ee
The spin connection is then expanded as
\be
\omega_{\mu ab} = \bar{\omega}_{\mu ab}
- \frac{1}{2} \left( \bar{\nabla}_a h_{\mu b}
- \bar{\nabla}_b h_{\mu a} \right)
+ {\cal O}(h^2)\ .
\label{om}
\ee
Using \eq{ads} and \eq{om}, and the field equation
$C_{ab}=-\mu(R_{ab}+2m^2\eta_{ab})$, up to quadratic terms in $h_{ab}$
we find
\bea
\slashchar{\widetilde\nabla}^L
&\equiv& \gamma^i {\widetilde\nabla}^L_i
= \slashchar{\bar{\hat\nabla}}^L
+ \Delta \slashchar{\widetilde\nabla}^L \ ,
\nn\w2
\Delta \slashchar{\widetilde\nabla}^L
&=& - \frac{1}{2} \c^i h_i{}^a \bar{\nabla}_a
- \frac{1}{4} \c^i \c^a Y_{i a}\ ,
\label{X}
\eea
where
\be
Y_{ab} \equiv \eps_b{}^{cd} \bar{\nabla}_c h_{ad}
+\frac{1}{\mu+m}  \left({\bar\nabla}^2 h_{ab}
+{\bar\nabla}_a {\bar\nabla}_b h
-2 {\bar\nabla}^c {\bar\nabla}_{(a} h_{b)c}
- 4m^2 h_{ab} \right) \ .
\label{Y}
\ee

To obtain the solution of $\slashchar{\widetilde\nabla}^L \eps=0$
we first note that the solution of
$\slashchar{\bar{\hat\nabla}}^L \epsilon_{AdS} = 0$ is
\be
\epsilon_{AdS} = e^{\frac{1}{2}\rho} \epsilon_0(\tau, \phi)
+ e^{-\frac{1}{2}\rho} \epsilon_1(\tau, \phi)\ ,
\label{adsws}
\ee
where $\epsilon_0$, $\epsilon_1$ are spinors satisfying
\be
\gamma_2 \epsilon_0 = \epsilon_0\ , \qquad
\partial_\phi^2 \epsilon_0 = - \frac{1}{4} \epsilon_0\ , \qquad
\epsilon_1 = 2 \gamma^1 \partial_\phi \epsilon_0\ .
\label{lkcond}
\ee
Writing the solution of $\slashchar{\widetilde\nabla}^L \eps=0$ as
$\epsilon = \epsilon_{AdS} + \Delta \epsilon$ we find
\be
\Delta \slashchar{\widetilde\nabla}^L \epsilon_{AdS}
+ \left( \slashchar{\bar{\hat\nabla}}^L
+ \Delta \slashchar{\widetilde\nabla}^L \right) \Delta \epsilon = 0\ .
\ee
For the HMT boundary conditions (\ref{nbc}) asymptotic behaviors
of $Y_{ab}$ in \eq{Y} 
are
\bea
Y_{--} &=& {\cal O}(e^{-(1+\mu\ell)\rho}) + {\cal O}(e^{-2\rho})\ ,
\nn\w2
Y_{-2},\ Y_{2-} &=& {\cal O}(e^{-(2+\mu\ell)\rho})
+ {\cal O}(e^{-2\rho})\ ,
\nn\w2
\mbox{other } Y_{ab} &=& {\cal O}(e^{-(3+\mu\ell)\rho})
+ {\cal O}(e^{-2\rho})\ .
\label{xasympt}
\eea
The asymptotic behaviors for the chiral point $\mu=m$ are given by
replacing $e^{(1-\mu\ell)\rho}$ with $\rho$ in \eq{xasympt}.
For the standard Brown-Henneaux boundary conditions \eq{sbc} there
are no $e^{-\mu\ell\rho}$ dependent terms in \eq{xasympt}.
Using \eq{X}, \eq{xasympt}, and $\gamma_2 \epsilon_0 = \epsilon_0$,
$\gamma_2 \epsilon_1 = -\epsilon_1$, we find
\bea
\Delta\slashchar{\widetilde\nabla}^L \left[
e^{\frac{1}{2}\rho} \epsilon_0 \right]
&=& \left[ {\cal O}(e^{-\left(\frac{3}{2}+\mu\ell\right)\rho})
+ {\cal O}(e^{-\frac{3}{2}\rho}) \right] \epsilon_0,
\nn\w2
\Delta\slashchar{\widetilde\nabla}^L \left[
e^{-\frac{1}{2}\rho} \epsilon_1 \right]
&=& \left[ {\cal O}(e^{-\left(\frac{3}{2}+\mu\ell\right)\rho})
+ {\cal O}(e^{-\frac{5}{2}\rho}) \right] \epsilon_1.
\label{deltanabla}
\eea
{}From \eq{deltanabla} and
$\slashchar{\bar{\hat{\nabla}}}^L = {\cal O}(1)$,
we find that the generalized Witten spinor
satisfying $\slashchar{\widetilde\nabla}^L \eps=0$ has an expansion
\be
\epsilon = e^{\frac{1}{2}\rho} \, \epsilon_0(\tau, \phi)
+ e^{-\frac{1}{2}\rho} \, \epsilon_1(\tau, \phi)
+ e^{-\left(\frac{3}{2}+\mu\ell\right)\rho} \, \tilde{\epsilon}_2(\tau, \phi)
+ e^{-\frac{3}{2}\rho} \, \epsilon_2(\tau, \phi)
+ \cdots\ .
\label{lexp}
\ee
The coefficient spinors $\epsilon_2$, $\tilde{\epsilon}_2,\ \cdots$
can be determined by solving $\slashchar{\widetilde\nabla}^L \eps=0$
iteratively, but we shall only need $\eps_0$ and $\eps_1$.

Similarly, we find that the generalized Witten spinor satisfying
$\slashchar{\widetilde\nabla}^R \epsilon = 0$ has an expansion
\be
\epsilon = e^{\frac{1}{2}\rho} \, \epsilon_0(\tau, \phi)
+ e^{-\frac{1}{2}\rho} \, \epsilon_1(\tau, \phi)
+ e^{-\left(\frac{1}{2}+\mu\ell\right)\rho} \, \tilde{\epsilon}_2(\tau, \phi)
+ e^{-\frac{3}{2}\rho} \, \epsilon_2(\tau, \phi)
+ \cdots\ ,
\label{rexp}
\ee
where $\epsilon_0$, $\epsilon_1$ are spinors satisfying
\be
\gamma_2 \epsilon_0 = - \epsilon_0\ , \qquad
\partial_\phi^2 \epsilon_0 = - \frac{1}{4} \epsilon_0\ ,\qquad
\epsilon_1 = - 2 \gamma^1 \partial_\phi \epsilon_0\ .
\label{rkcond}
\ee
To obtain the expansion \eq{rexp} we have used
\bea
\Delta\slashchar{\widetilde\nabla}^R \left[
e^{\frac{1}{2}\rho} \, \epsilon_0 \right]
&=& \left[ {\cal O}(e^{-(\frac{1}{2}+\mu\ell)\rho})
+ {\cal O}(e^{-\frac{3}{2}\rho}) \right] \epsilon_0,
\nn\w2
\Delta\slashchar{\widetilde\nabla}^R \left[
e^{-\frac{1}{2}\rho} \, \epsilon_1 \right]
&=& \left[ {\cal O}(e^{-(\frac{5}{2}+\mu\ell)\rho})
+ {\cal O}( e^{-\frac{5}{2}\rho}) \right] \epsilon_1,
\eea
which have different behaviors from \eq{deltanabla} due to
the different chiralities of $\epsilon_0$, $\epsilon_1$.

Finally we observe that upon choosing
\be
\epsilon_0 = \cos\ft12 (\tau+\phi) \, \eta_+^1
+ \sin\ft12 (\tau+\phi) \, \eta_+^2\ ,
\label{thec}
\ee
where $\eta_+^A$ ($A=1,2$) are arbitrary constant Majorana spinors
satisfying $\gamma_2 \eta_+^A = \eta_+^A$, determining $\epsilon_1$
by \eq{lkcond}, and using the results of Appendix \ref{app:killing},
we find that $\epsilon$ approaches the Killing spinor
$\epsilon_K=\eps^{1}_K+\eps^{2}_K$ in \eq{kspinor}
\be
\epsilon = \epsilon_K
+ {\cal O}(e^{-\left(\frac{3}{2}+\mu\ell\right)\rho})
+ {\cal O}(e^{-\frac{3}{2}\rho})\ .
\ee
We can also choose $\eps_0$ as in \eq{thec} by replacing $\phi$
by $-\phi$, and $\eta_+^A$ by $\eta_-^{\dot{A}}$ satisfying
$\gamma_2 \eta_-^{\dot{A}} = - \eta_-^{\dot{A}}$,
and determine $\epsilon_1$ by \eq{rkcond}. This results in
Killing spinors $\eps_K^{\dot 1} + \eps_K^{\dot 2}$ given
in \eq{rc}. These are the choices we shall make for $\eps_0$ in the
remainder of this paper. In this way, the Killing vectors
of $SO(2,2)$ emerge as bilinears of appropriate Killing spinors
(see Appendix \ref{app:killing}) in the computation of the
Witten-Nester charges as boundary integrals.


\section{ The Abbott-Deser-Tekin Charges }


\subsection{The Relation Between the Witten-Nester and ADT Charges}


Having determined the asymptotic solution for the generalized Witten spinor we can now study the boundary integral \eq{ab} for the Witten-Nester charges. We shall employ the HMT boundary conditions for different values of $\mu$ and compare the result with the ADT charges reviewed in Appendix \ref{app:adt}.

Let us first examine the second term in the formula \eq{ab} for the Witten-Nester charges $Q^L_{WN}$. Using the asymptotic solution for the Witten spinor described in the previous section,  and the notation (see Appendix (\ref{app:killing}))
\be
\epsilon_1 \rightarrow \eps^{B}\ ,   \qquad
\epsilon_2 \rightarrow \eps^{A}\ ,
\label{limits}
\ee
for the asymptotic values of the Witten spinors, with $A,B=1,2$ labeling the Killing spinors as described in Appendix \ref{app:killing}, and using the definition of the Killing vectors $\xi^{AB}$ in terms of these spinors as defined in \eq{k1}, we readily find that
\bea
\frac{2}{\mu} \int d\phi \, \sqrt{-g} \,
{\bar\eps}^A \c^\sigma\eps^B \left( \cG_{\phi\sigma}
- \textstyle{\frac12} g_{\phi\sigma} \cG \right)
&=& \frac{2}{\mu}  \int d\phi \, \sqrt{-\bar{g}} \,
\xi^{AB\sigma} \left( \delta\cG_{\phi\sigma}
-\ft12 \bar{g}_{\phi\sigma} \delta\cG\right)
\label{aterm} \w2
&=& \frac2{\mu} \int d\phi \, \sqrt{-\bar{g}} \,
f_C^{\tau\rho}(\xi)\ ,
\label{2ndterm}
\eea
where $f_C^{\mu\nu}$ is defined in \eq{line1}, and $\delta \cG$ denotes
the linearized $\cG$. In \eq{aterm}, we have made use of the fact that
at the boundary of $\Sigma$, which is at radial infinity
$\rho \rightarrow \infty$, the spinors approach their Killing spinor values
and that in the remaining part of the integrand only the contributions
that are linear in the deviation field survive. Comparing this result
with the last term in the ADT charge formula \eq{dt}, we see that they
are in agreement up to a normalization factor of $16\pi G$, which has
been set to $16\pi G = 1$ in the calculations of the Witten-Nester charges.
Thus, it remains to show that the first term in \eq{ab} produces
the first term in \eq{dt}, again up to an overall normalization factor
of $16\pi G$. Letting $\eps_1 \rightarrow \eps^B$, and
$\eps_2 \rightarrow \eps^A$ (see Appendix \ref{app:killing}), and
using the expansion \eq{lexp} with the asymptotic solution as given
in Section 4.2, we compute
\bea
\varepsilon^{\tau\rho\phi} \,
\bar{\epsilon}^A \hat{\nabla}_\phi \epsilon^B
&=& \frac{1}{\sqrt{-g}}
\left[ e^{\frac{1}{2}\rho} \, \bar{\epsilon}^A_0 + \cdots \right]
\left[ \bar{\hat{\nabla}}_\phi
- \frac{1}{4} \bar{e}_\phi{}^1 \left( \bar{\nabla}_a h_{1b} \gamma^{ab}
+ m h_{1a} \gamma^a \right) + {\cal O}(h^2) \right]
\nn\w2
&& \times \left[ e^{\frac{1}{2}\rho} \epsilon^B_0
+ e^{-\frac{1}{2}\rho} \epsilon^B_1 + \cdots \right]
\nn\w2
&=& - \frac{1}{2\ell^2} \, \bar{\epsilon}^A_0
\left( \bar{\nabla}_a h_{1b} \gamma^{ab} + mh_{1a} \gamma^a \right)
\epsilon^B_0 + \cdots
\nn\w2
&=& \frac{1}{2\ell^2} \, \bar{\epsilon}^A_0 \gamma^+ \epsilon^B_0
\left( \bar{\nabla}_2 h_{+1} - \bar{\nabla}_+ h_{21}
- mh_{+1} \right) + \cdots\ ,
\label{nwresult}
\eea
where $\cdots$ denote higher order terms in $e^{-\rho}$.
To compare this result with the first term in \eq{dt}, we compute
\be
\cF^{\tau\rho}_E(\xi) = \frac{1}{\ell^2} e^{-\rho} \xi^{AB\,+} \left(
\bar{\nabla}_2 h_{+1} - \bar{\nabla}_+ h_{21} - m h_{+1} \right)
+ \cdots\ ,
\ee
where we have used the fact $\xi^{AB\,+} = {\cal O}(e^\rho)$,
$\xi^{AB\,-} = {\cal O}(e^{-\rho})$, $\xi^{AB\,2} = {\cal O}(1)$.
Thus, the first term in \eq{ab}, together with the second term computed
in \eq{2ndterm} evaluated for the Killing vectors $\xi^{AB\mu}$,
which we denote by $\xi^{(+)\mu}$ for simplicity, and give in \eq{k1},
yield the total
\be
Q_{WN}^L[\xi^{(+)}]
= \int d\phi \sqrt{-\bar{g}} \left[ 2\left(1+\frac{m}{\mu}\right)
\cF^{\tau\rho}_E(\xi^{(+)})
+  \frac2{\mu} f_C^{\tau\rho} (\xi^{(+)}) \right]
 .
 \label{56}
\ee
Comparing with the ADT charges described in Appendix \ref{app:adt}, and
reintroducing the factor of $16\pi G$ which has been set equal to one
in \eq{56}, we see that we have established the relation
\be
Q_{WN}^L[\xi^{(+)}] = Q_{ADT}[\xi^{(+)}]\ .
\label{ag}
\ee
The Killing vector $-2\ell^{-1}K_0 = \ft12 (\xi^{11} + \xi^{22})$
in \eq{kn} corresponds to the charge $E-mJ$, where $E$ is the energy
and $J$ is the angular momentum. Similarly, taking for the Witten spinors
$\eps^{\dot A}$ and $\eps^{\dot B}$, we find that $Q_{WN}^R$ (related
to $Q_{WN}^L$ by letting $m \rightarrow -m$) becomes
\be
Q_{WN}^R[\xi^{(-)}] = Q_{ADT}[\xi^{(-)}]\ ,
\label{ag2}
\ee
where $\xi^{(-)}$ denote the Killing vectors $\xi^{\dot{A}\dot{B}}$
in \eq{k2}. The Killing vector $-2\ell^{-1} J_0
= \ft12 (\xi^{\dot{1}\dot{1}} + \xi^{\dot{2}\dot{2}})$
in \eq{kn} corresponds to the charge $E+mJ$.


\subsection{ Computation of the ADT Charges with HMT Boundary Conditions}


Having established the relations \eq{ag} and \eq{ag2} between the
Witten-Nester charges and the ADT charges, in this section we shall
compute these charges for the HMT boundary conditions, to ensure
that they are actually finite.

The conserved charges we shall compute are within the $SO(2,2)$
subalgebra of two copies of Virasoro algebra that have been shown to
arise as the asymptotic symmetry algebra \cite{Henneaux:2009pw}.
Thus, we compute (see Appendix \ref{app:conv} for notation)
\be
Q_{ADT}[\xi^{(\pm)}]
= \frac{1}{8\pi G} \int d\phi \sqrt{-\bar{g}} \left[
\left( 1 \pm \frac{m}{\mu} \right) {\cal F}_E^{\tau\rho}(\xi^{(\pm)})
+ \frac{1}{\mu} f_C^{\tau\rho}(\xi^{(\pm)}) \right]\ .
\label{01}
\ee
For the first terms of the integrands, using the definition of
$\cF_E^{\mu\nu}$ given in \eq{fe}, the boundary conditions \eq{nbc},
which also cover the case \eq{sbc} for $|\mu\ell|>1$ by simply setting
the $\tf$ field to zero, and \eq{ftrace}, we obtain
\bea
{\cal F}_E^{\tau\rho} (\xi^{(+)})
&=& - \sqrt{2} \, m^3 e^{-3\rho} \, \xi^{(+)+} f_{++} + \cdots,
\nn\w2
{\cal F}_E^{\tau\rho} (\xi^{(-)})
&=& \sqrt{2} \, m^3 e^{-3\rho} \, \xi^{(-)-} \left[ f_{--} +
\ft12 (1+\mu\ell) e^{(1-\mu\ell)\rho} \tilde{f}_{--} \right] + \cdots\ ,
\label{03}
\eea
where $\cdots$ are higher order terms which do not contribute to the
charge. When substituted into the charge, the $e^{(1-\mu\ell)\rho}$
term is divergent for $\rho \rightarrow \infty$.

The second terms of the integrands, upon using \eq{line2}, \eq{lincg},
and the boundary conditions \eq{nbc}, yield the results
\bea
f_C^{\tau\rho}(\xi^{(+)})
&=& 
\cdots,
\nn\w2
f_C^{\tau\rho}(\xi^{(-)})
&=& \frac{1}{\sqrt{2}} m^4 (1-\mu^2\ell^2) e^{-(2+\mu\ell)\rho} \,
\xi^{(-)-} \tilde{f}_{--} + \cdots \ ,
\label{04}
\eea
where, again, $\cdots$ are higher order terms which do not
contribute to the charge.  Substituting \eq{03} and \eq{04} into \eq{01}
we obtain the ADT charges as
\be
Q_{ADT}[\xi^{(+)}]
= - \frac{1}{16\sqrt{2} \pi G} \int d\phi
\left( 1 + \frac{m}{\mu} \right) e^{-\rho} \xi^{(+)+} f_{++}\ ,
\qquad {\rm for\ } \mu\ell \ge 1\ ,
\label{05}
\ee
and
\be
Q_{ADT}[\xi^{(-)}]  = \left\{
\begin{array}{ll}
\displaystyle{
\frac{1}{16\sqrt{2} \pi G} \int d\phi
\left( 1 - \frac{m}{\mu} \right) e^{-\rho} \xi^{(-)-} f_{--}
}\ ,
& \qquad {\rm for}\  \mu \ell > 1\ , \\
&\\
\displaystyle{
\frac{1}{16\sqrt{2} \pi G} \int d\phi \, e^{-\rho}
\xi^{(-)-} \tilde{f}_{--}}\ ,
& \qquad {\rm for}\  \mu\ell =1 \ .
\end{array}
\right.
\label{cases}
\ee

\bigskip

\noindent In the computation of the second equation in \eq{cases}, the
divergent terms proportional to $\tilde{f}$ in ${\cal F}_E$ and $f_C$
have canceled out.  We see that while one may expect
$Q_{ADT}[\xi^{(-)}]$ to vanish at the chiral point, in fact it does
not, as can be seen from the
implementation of the prescription \eq{rule}. Thus, it is remarkable
that even at the chiral point $\mu\ell=1$, both charges associated
with $\xi^{(+)}$ and $\xi^{(-)}$ are nonvanishing.
This is in accordance with what has been recently found in
\cite{Henneaux:2009pw} by apparently different methods.

In the case of $\mu\ell=1$ and the {\it standard Brown-Henneaux boundary
conditions} \eq{cbc}  with the $\tf$ terms set to zero, the ADT charges
are given by
\be
\mu\ell=1 : \qquad \left\{
  \begin{array}{ll}
   \displaystyle{Q_{ADT}[\xi^{(+)}]
= - \frac{1}{8\sqrt{2} \pi G} \int d\phi \, e^{-\rho}
\xi^{(+)+} f_{++}}\ , &  \\
   \displaystyle{Q_{ADT}[\xi^{(-)}] = 0}\ . &
  \end{array}
\right.
\label{ct1}
\ee

\noindent
Therefore, in order to prove a positive energy theorem for the theory at
the chiral point and obeying the standard Brown-Henneaux boundary
conditions, it must be shown that for any solution of the theory, the
above expression for $Q_{ADT}[\xi^{(+)}]$ is positive. While we have not proven such a theorem here, we have established the bound \eq{mpc}.

At present the only known exact solutions that satisfy the standard Brown-Henneaux boundary conditions have vanishing Cotton tensor, and consequently they are conformal
to AdS. Normalizing the energy of AdS to be vanishing, all the remaining solutions
have either conical singularity for $GM <0$ and therefore excluded from the physical spectrum, or they have positive energy. What we do not know at present is whether there exist other solutions which violate the bound \eq{mpc}. As such, a positive energy theorem based on \eq{mpc} is not ruled out.

While the ADT energy for the solutions with vanishing Cotton tensor mentioned above can be directly computed from the ADT charge formula, and therefore the Witten-Nester bulk integral is not needed, it is useful to compare the ADT result with the consequences of the bound \eq{mpc}. We shall do this analysis in the next section for the general solution with vanishing Cotton tensor as well as the pp-waves in AdS which satisfy the weaker version of the Brown-Henneaux boundary conditions for completeness. As a result, we will see that the need for the existence of a globally well defined solution of the generalized Witten spinor equation \eq{ws2}.


\subsection{The General Solution With Vanishing Cotton Tensor }


The most general solution of CTMG with vanishing Cotton tensor is
locally AdS metric which can be written as \cite{Banados:1998gg}
\be
ds^2 = 2 \left( (mr)^2 + \frac{1}{4(mr)^2} h(u) \bar{h}(v) \right) du dv
+ \frac{dr^2}{(mr)^2} + h(u) du^2 + \bar{h}(v) dv^2\ ,
\label{banados}
\ee
where
\be
u = \frac{1}{\sqrt{2}} \left( - t + \ell \phi \right)\ , \qquad
v = \frac{1}{\sqrt{2}} \left( t + \ell \phi \right)\ ,
\ee
and $h, {\bar h}$ are real and independent arbitrary functions.
This solution can be obtained from the ``vacuum solution'' defined
by $h=0$, $\bar h=0$ by means of conformal transformations. For example,
starting from the vacuum solution, the following  transformations
\be
u \rightarrow f(u)\ , \qquad
v \rightarrow v-\frac1{4m^4r^2}\,\frac{f''}{f'}\ , \qquad
r \rightarrow \frac{r}{\sqrt {f'}}\ ,
\label{trans}
\ee
where $f'= \partial f / \partial u$ yield the metric \eq{banados} with
\be
h(u) = -\frac{1}{2m^2} \{f,u\}\ ,\qquad \bar h(v) =0\ ,
\ee
with the Schwarzian derivative defined as
\be
\{f,u\} = \left(\frac{f''}{f'}\right)'
- \frac12 \left(\frac{f''}{f'}\right)^2\ .
\ee
A more general coordinate transformation involving in particular
$v\rightarrow g(v)+\cdots$ which generates the $\bar h(v)$ dependent
terms in \eq{banados} as ${\bar h}(v)=-\frac1{2m^2} \{g,v\}$, is more
complicated but it will not be needed here.

Although the general metric \eq{banados} is locally AdS, it
represents physically distinct configurations for distinct values
of $h$ and $\bar h$ (see \cite{Banados:1998gg} for a review).
In particular, choosing $h$ and $\bar h$ to be constants as
\be
h= 4G (M_\infty + m J_\infty)\ ,\qquad
{\bar h}= 4G (M_\infty - m J_\infty)\ ,
\label{sv}
\ee
and redefining the radial coordinate as
\be
{\tilde r}^2= r^2 +\frac{4G^2\ell^2}{r^2}
\left(M_\infty^2\ell^2 -J_\infty^2 \right) + 4GM_\infty \ell^2\ ,
\ee
gives the BTZ black hole solution \cite{Banados:1992wn}
\be
ds^2= - \left(m^2 {\tilde r}^2 - 8GM_\infty \right) dt^2
- 8GJ_\infty dt d\phi +{\tilde r}^2 d\phi^2
+ \frac{d{\tilde r}^2}{m^2 {\tilde r}^2 - 8GM_\infty
+ \frac{16G^2J_\infty^2}{ {\tilde r}^2}}\ ,
\label{btz}
\ee
In the $\mu=\infty$ theory without the CS term the parameters $M_\infty$
and $J_\infty$ represent the mass and the angular momentum of the
black hole respectively.
The radial coordinate $\tilde{r}$ covers the exterior region of
the horizon $\tilde{r} \ge r_+$ when $r$ varies from
$r=\ft12 \sqrt{r_+^2-r_-^2}$ to $r=\infty$.
Here, $r_+$ and $r_-$ ($r_+ \ge r_-$) are roots of
$m^2 {\tilde r}^2 - 8GM_\infty + \frac{16G^2J_\infty^2}{ {\tilde r}^2}=0$,
and are related to $M_\infty$ and $J_\infty$ as
$r_+^2 + r_-^2 = 8GM_\infty\ell^2$,\ \ $r_+ r_- = 4GJ_\infty\ell$.
As noted in \cite{Banados:1992wn}, the ``vacuum metric'', which is
defined as the $J_\infty=0=M_\infty$ case, is not the AdS metric,
and the latter arises for $J_\infty=0$ and $GM_\infty=-1/8$.
Moreover, the solution has conical
singularity for any other negative value of $GM_\infty$.

For the general solution \eq{banados}, the deviations from the AdS
metric also obey the boundary conditions \eq{sbc} as can be seen from
\bea
h_{++} &=& ( \bar{h} + \ft12 ) \frac{1}{(mr)^2} + {\cal O}(r^{-4})\ ,
\nn\w2
h_{--} &=& ( h + \ft12 ) \frac{1}{(mr)^2} + {\cal O}(r^{-4})\ ,
\nn\w2
h_{+-} &=& -\frac{1}{2(mr)^2} + {\cal O}(r^{-4})\ ,
\nn\w2
h_{22} &=& \frac{1}{(mr)^2} + {\cal O}(r^{-4})\ ,
\eea
and the relation $mr = \sinh\rho = \ft12 e^{\rho} + {\cal O}(e^{-\rho})$.
Thus, the ADT charges are obtained from \eq{05} and \eq{cases} as
\be
Q_{ADT}[\xi^{(+)}]
= -\frac{1}{8\sqrt{2} \pi G} \int_0^{2\pi} d\phi
\left( 1 + \frac{m}{\mu} \right) e^{-\rho} \xi^{(+)+} ( 2\bar h+1) \ ,
\qquad {\rm for\ } \mu\ell \ge 1\ ,
\label{05b}
\ee
and
\be
Q_{ADT}[\xi^{(-)}]  = \left\{
\begin{array}{ll}
\displaystyle{
\frac{1}{8\sqrt{2} \pi G} \int_0^{2\pi} d\phi
\left( 1 - \frac{m}{\mu} \right) e^{-\rho} \xi^{(-)-} (2 h +1)
}\ ,
& \qquad {\rm for}\  \mu \ell > 1\ , \\
&\\ 0\ ,
& \qquad {\rm for}\  \mu\ell =1 \ ,
\end{array}
\right.
\label{casesb}
\ee
where $\xi^{(+)}$ and $\xi^{(-)}$ are Killing vectors in \eq{k1}
and \eq{k2} respectively.
For the special values \eq{sv} of $(h,\bar h)$ corresponding to the BTZ
black hole, the ADT charges $E \mp mJ$ corresponding to the Killing
vectors $-2\ell^{-1}K_0$ and $-2\ell^{-1}J_0$ in \eq{kn} are thus
\be
E - mJ =  \left( 1 + \frac{m}{\mu} \right) \left(M_\infty +
\frac{1}{8G} -mJ_\infty \right) \ ,
\qquad {\rm for\ } \mu\ell \ge 1\ ,
\label{05c}
\ee
and
\be
E + mJ = \left\{
\begin{array}{ll}
\displaystyle{
\left( 1 -\frac{m}{\mu} \right) \left(M_\infty
+ \frac{1}{8G} +mJ_\infty \right)} \ ,
& \qquad {\rm for\ } \mu \ell > 1\ , \\
&\\ 0\ ,
& \qquad {\rm for\ } \mu\ell =1 \ .
\end{array}
\right.
\label{casesc}
\ee

These results suggest that the ADT charges can be positive or negative depending on the choice for $h(u)$ and $\bar{h}(v)$. On the other hand, we have the bounds \eq{mpc} and \eq{mp4}, where the vanishing of $X_{\mu\nu}$ for the solutions \eq{banados} implies that
$Q_{WN}[\xi^{(\pm)}] \ge 0$, and in view of \eq{ag} and \eq{ag2} it means
that $Q_{ADT}[\xi^{(\pm)}]$ must be positive. The resolution of this
apparent contradiction lies in the fact that the bounds \eq{mpc} and
\eq{mp4} assume the existence of globally well defined solutions
of the generalized Witten equations, in this case for $\mu\ell > 1$.
It follows that such solutions must fail to exist for the choices of
$h$ and $\bar{h}$ which yield a negative value for $Q_{ADT}[\xi^{(\pm)}]$.
To see this we proceed as follows. We  substitute into
$\gamma^i {\widetilde\nabla}^R_i \eps=0$ (see \eq{ws2}
and the discussion above \eq{mp4})
\be
\epsilon = \epsilon_0^R + \Delta\epsilon\ ,
\ee
where $\eps_0^R$ is the Killing spinor of the ``vacuum solution''
\eq{banados} with $h=\bar{h}=0$ that satisfies
$( \nabla_\mu + \frac{1}{2} m\gamma_\mu )^{\rm vac} \epsilon_0^R = 0$,
and is given by
\be
\epsilon_0^R = (mr)^{\frac{1}{2}} \left( \eta_-
- m \, u \, \gamma_- \eta_+ \right)
+ (mr)^{-\frac{1}{2}} \eta_+
\ee
in terms of arbitrary constant spinors $\eta_\pm$ satisfying
$\gamma_2 \eta_\pm = \pm \eta_\pm$. Next, we use the solution
\eq{banados} with $\bar h=0$, and choose the following dreibein
\be
e_v{}^+ = mr\ , \qquad
e_u{}^+ = \frac{h}{2mr}\ , \qquad
e_u{}^- = mr\ , \qquad
e_r{}^2 = \frac{1}{mr}\ .
\ee
In this basis the only non-vanishing components of the spin connection are
\be
\omega_v{}^{+2} = m^2r\ , \qquad
\omega_u{}^{+2} = r \partial_r \left( \frac{h}{2r} \right)\ , \qquad
\omega_u{}^{-2} = m^2r\ .
\ee
Next, we use the fact that the Cotton tensor vanishes for this solution,
and in the equation $\gamma^i {\widetilde\nabla}^R_i \eps=0$
we project to the $\gamma_2 = \pm 1$ subspaces and substitute the expansions
\be
\Delta\epsilon_- =  \sum_{n=1}^\infty (mr)^{-2n + \frac{1}{2}}
\epsilon_{n-}\ ,\qquad
\Delta\epsilon_+ = \sum_{n=1}^\infty (mr)^{-2n - \frac{1}{2}}
\epsilon_{n+}\ .
\label{epsexp}
\ee
Thus we find recursion relations
\bea
\epsilon_{n+} &=& \frac{1}{2(2n-1)m} \gamma_+ \partial_u \epsilon_{n-}\ ,
\label{r1}\w2
\epsilon_{1-} &=& - \frac{1}{4} h
\left( \eta_- - m \, u \, \gamma_- \eta_+ \right)\ ,
\label{r2}\w2
2(n+1) \epsilon_{n+1\, -}
&+& \frac{1}{2m} \gamma_- \partial_u \epsilon_{n+}
+ \frac{1}{2} h \epsilon_{n-} = 0\ ,
\label{9}
\eea
To make further progress, at this point we consider the case $h=$
constant. In this case we can solve the recursion relations and the
using this result in \eq{epsexp}, we find
\be
\epsilon_-
= (mr)^{\frac{1}{2}} e^{-\frac{h}{4(mr)^2}}
\left( \eta_- - m \, u \, \gamma_- \eta_+ \right)\ .
\ee
We now see that in the limit $r \rightarrow 0$, $\Delta\epsilon_-$ is
divergent for $h < 0$, while it is finite for $h > 0$. This suggests
that a globally defined Witten spinor does not exist for $h < 0$.


\subsection{The Chiral pp-Waves}


The chiral pp-wave solutions are exact solutions of CTMG which
take the form \cite{AyonBeato:2004fq,Carlip:2008jk,gps}
\be
ds^2 = 2 (mr)^2 du dv + \frac{dr^2}{(mr)^2} + h(u, r)\, du^2\ ,
\label{gps}
\ee
where, using the notation $h(u,r)\equiv h$, we have
\bea
\mu\ell \not= \pm 1: \ \
h &=& (mr)^{1-\mu\ell} f_1(u) + (mr)^2 f_2(u) + f_3(u)\ ,
\label{ss1}\w2
\mu\ell = +1: \ \
h &=& \ln (mr) \, f_1(u) + (mr)^2 f_2(u) + f_3(u)\ ,
\label{s2}\w2
\mu\ell = -1: \ \
h &=& (mr)^2 \ln(mr) \, f_1(u) + (mr)^2 f_2(u) + f_3(u)\ ,
\label{s3}
\eea
and $f_1,f_2,f_3$ are arbitrary functions of $u$. While $f_2(u)$ and $f_3(u)$ can be removed by local coordinate transformations, the global nature of the solution may depend on them. Of these solutions, the ones which satisfy the HMT boundary conditions are
\bea
|\mu\ell| > 1\ ,\ \mu=-1 : \ \ h &=& f_3(u)\ ,
\label{sss1}\w2
0< |\mu\ell| <1: \ \ h &=& (mr)^{1-\mu\ell} f_1(u) + f_3(u)\ ,
\label{sss2}\w2
\mu\ell = +1: \ \ h &=& \ln (mr) \, f_1(u)  + f_3(u)\ .
\label{sss3}
\eea
The first case corresponds to the solution \eq{banados} with $\bar h=0$ and $h=f_3$ which we have already dealt with in the previous section, and therefore in what follows we shall consider the remaining cases only.

Applying the formula \eq{05}, \eq{cases} for the charges of these
solutions we find
\footnote{These charges were computed for constant $f$'s in
\cite{Olmez:2005by}, and for these cases our results agree except the
second equation in \eq{cases2}.}
\be
 Q_{ADT}[\xi^{(+)}] = -\frac{1}{8\sqrt{2} \pi G} \int d\phi \,
\left( 1 + \frac{m}{\mu} \right) e^{-\rho} \xi^{(+)+}
\label{result}
\ee
for all values of $\mu$, which is independent of $f_3$, $f_1$
and {\it positive}, and
\be
Q_{ADT}[\xi^{(-)}]  = \left\{
\begin{array}{ll}
\displaystyle{\frac{1}{8\sqrt{2} \pi G} \int d\phi \,
\left( 1 - \frac{m}{\mu} \right) e^{-\rho} \xi^{(-)-}
\left( 2f_3(u) + 1 \right)}\ ,
& \quad {\rm for}\  \ 0 < |\mu \ell| < 1\ , \\
&\\
\displaystyle{\frac{1}{8\sqrt{2} \pi G} \int d\phi \, e^{-\phi}
\xi^{(-)-} \, 2f_1(u)}\ ,
& \quad {\rm for}\ \ \mu\ell =1 \ .
\end{array}
\right.
\label{cases2}
\ee

\bigskip

\noindent Thus for the chiral models, imposing the {\it standard
Brown-Henneaux boundary conditions}, namely \eq{cbc} with the $\tf$
fields set to zero, gives vanishing ADT charge for $\xi^{(-)}$:
\be
\mu\ell=1: \qquad
Q_{ADT}[\xi^{(-)}]=0\ .
\label{2c}
\ee
This result is consistent with the fact that the extreme BTZ black hole
solution which is obtained from \eq{btz} by setting $M_\infty=mJ_\infty$
\cite{Behrndt:1999jp} has vanishing
$Q_{ADT}[\xi^{(-)}]$.


\section{ Comments}


The bounds \eq{mpc} and \eq{mp4} we have established on the
Witten-Nester energy highlight the importance of the Cotton tensor.
Therefore it is useful to examine the classification of $3D$ spacetimes
based on the eigenvalue problem for the Cotton tensor,
$ (C^a{}_b -\lambda\delta^a_b) V^b=0$, where $\lambda\subset C$.
Such a classification is available \cite{Hall:1987bz,Hall:1999}
(see also \cite{Torres:2003,Garcia:2003bw,Sousa:2007ax} ), and it
shows that the possible canonical forms of the Cotton tensor are
as follows \cite{Sousa:2007ax,Garcia:2003bw} %
\footnote{The labeling in the first column is motivated by the Petrov
classification of $4D$ spacetimes.}:

\be
\begin{array}{cl}
           ${\rm Petrov\ Type}$  &  \qquad ${\rm Canonical\ Form}$   \\
             N &   \qquad C_{ab}= \lambda k_a k_b \\
               D &   \qquad C_{ab}= \alpha (\eta_{ab} -3 m_a m_b )\\
             D' &  \qquad C_{ab}= \alpha (\eta_{ab} +3t_a t_b )\\
            III &   \qquad C_{ab}= 2 \tau  k_{(a} m_{b)} \\
              II &   \qquad C_{ab}= \alpha (\eta_{ab} -3 m_am_b) +\lambda k_a k_b \\
              I &    \qquad C_{ab}= \alpha (\eta_{ab} -3 m_am_b) -\beta\, (k_a k_b +n_a n_b ) \\
              I' &  \qquad C_{ab}= \alpha (\eta_{ab} -3 m_am_b) -\beta\, (k_a k_b -n_a n_b ) \\
\end{array}
\ee

\bigskip

\noindent where $\alpha,\beta \subset R$ and $\beta \ne 0$, and it is possible to choose $\lambda=\pm 1$ and $\tau=\pm 1$. Furthermore, $k^a$ and $n^a$ are null vectors and $m^a$ is a spacelike vector, with the only nonvanishing inner products $k^a n_a=-1$, and $m^a m_a=1$. We also have the timelike vector $t^a=(k^a+n^a)/{\sqrt 2}$ and spacelike vector $z^a=(k^a-n^a)/{\sqrt 2}$.

Assuming that the Witten-Nester energy does not depend on the choice of initial spacelike surface, we can choose this surface such that $u^a = t^a$. Then we find
\be
\begin{array}{cll}
        ${Petrov\ Type}$      &    \qquad - X_{ab} t^a v^b\\
            N &  \qquad  0\\
            D & \qquad 2\alpha^2 t^a v_a \\
            D' & \qquad -\alpha^2 t^a v_a \\
            III & \qquad \frac1{\sqrt 2} \tau^2 k^a v_a  \\
             II &   \qquad 2\alpha^2 t^a v_a +{\sqrt 2}\lambda \alpha\, k^a v_a\\
              I &     \qquad 2\alpha(\alpha-\beta) t^a v_a \\
              I' &  \qquad 2\alpha^2 t^a v_a - 2\alpha\beta\, z^a v_a \\
\end{array}
\ee
The vector $v^\mu$ is bilinear in spinors that obey the generalized Witten equation and asymptote the suitable Killing spinors. From these results, we see that there is no evidence for the positivity of Witten-Nester energy, with the exception of Type $N$ spacetimes, and Type $I$ spacetimes with $\alpha=\beta$. In the latter case, $X_{ab}$ does not vanish but $X_{ab} t^b=0$.

Given a Petrov-type, determining the corresponding space of solutions, and among the class those which obey the standard Brown-Henneaux boundary conditions is a notoriously difficult and so far unsolved problem. It is clear that all the solution of ordinary AdS gravity with the CS term absent are also the solutions of CTMG in which the Cotton tensor vanishes. All of these solutions are conformal to AdS, and they are well understood (see \cite{Banados:1998gg} for a review). Outside  this class, the only exact solutions of CTMG that have appeared in the literature until the present time are remarkably few and they are\footnote{A number of solutions that have appeared in the literature
have been shown in \cite{cps:2009} to be either a pp-wave or squashed
(stretched) AdS in disguise.} \cite{cps:2009}:

\begin{itemize}

  \item The general pp-wave solutions \cite{AyonBeato:2004fq,Carlip:2008jk,gps} are Type $N$ and they follow from the requirement of one null Killing vector \cite{gps}\footnote{The black hole solution of \cite{Garbarz:2008qn}, which obeys the weak version of the Brown-Henneaux boundary condition, is a coordinate transformation of the pp-wave at the chiral point \cite{AyonBeato:2004fq,Carlip:2008jk,gps} with a compactified spatial coordinate \cite{Garbarz:2008qn,cps:2009}.}.

  \item The timelike and spacelike squashed $(\mu\ell<3)$ or stretched $(\mu \ell>3)$ solutions admitting $SL(2,R) \times U(1)$ Killing vectors are of Type $D$ \cite{Nutku:1993eb,Gurses:1994}. These solutions have a squashing parameter related to $\mu$, and the discrete quotients are ``warped'' $AdS_3$ black holes \cite{Anninos:2008fx} which asymptote to squashed (stretched)  $AdS_3$.

\end{itemize}

Of these, only the Type $N$ solutions obey the standard Brown-Henneaux
boundary conditions, and a direct calculation that does not rely on the
Witten-Nester identity shows that their energy is positive. As mentioned
in the introduction, however, the bound \eq{mpc4} is nontrivial since we
do not know if all solutions of Type $N$ are necessarily chiral pp-waves.

Of the known Type $D$ solutions, the conserved charges for the  warped
AdS black holes have been computed in \cite{Bouchareb:2007yx}, where the
generalized version of the ADT formula is derived and used. The mass
turns out to be positive for $|\mu\ell|>3$ for the solution considered
in \cite{Bouchareb:2007yx} which is related to that of
\cite{Nutku:1993eb,Gurses:1994} by the double Wick rotation
$t\rightarrow it, \phi \rightarrow i\phi$. In  comparing this result
with the bound on the energy we have derived here, we note that while
in showing the equivalence of the Witten-Nester charge with the ADT
charge we assumed the asymptotically AdS HMT boundary conditions,
passing over to bulk integral by means of Stokes' theorem does not
depend on this assumption. It rather depends on the existence of a
globally well defined solution of the generalized Witten equation. Thus
the bounds \eq{mpc} and \eq{mp4} are valid for any solution of the TMG
equations of motion, not necessarily asymptoting to $AdS$, provided that
the global Witten spinors exist. However, the relation between the
Witten-Nester charges and ADT charges in presence of squashed AdS
boundary conditions needs to be established in this case, before a
rigorous comparison with the direct calculation of the mass described above.

The Witten-Nester identity we have found relies on a supercurrent
associated with local supersymmetry of topologically massive
supergravity in first order formulation. Since Noether currents
associated with local symmetries are defined up to divergence of an
antisymmetric tensor, one may consider an alternative definition of
energy which may lead to a positive energy theorem. Note, however, that
the charge definition we have used here, which is due to \cite{silva}
and generalizes the Hamiltonian approach of \cite{Regge:1974zd} by
employing superpotentials, passes important tests. In particular, it
produces the appropriate  conserved quantities of the BTZ black hole
that take into account the dependence on the CS coupling constant
\cite{Deser:2003vh,Olmez:2005by}. The charges constructed in this way
must also produce the appropriate charge algebra. As explained in
\cite{silva}, the first order formulation, which we have used here as
well, and the antisymmetry of the superpotential, such as the one we
have given in \eq{u} for CTMG, play crucial role in achieving these
properties. Indeed, we have found that the Witten-Nester charge defined
from the supertransformation of the supercurrent coincide with the ADT
charges $H \pm P$ (for null Killing vectors), which verifies the
anticommutation relation of supercharges $\{ Q, Q \} = H \pm P$.
Furthermore, the procedure of \cite{silva} for constructing the
conserved charges has been tested successfully in many models.

The Witten-Nester identity relies also on the existence of regular solutions to a generalized version of the Witten equation on the spatial slice. In Section 4 we showed that one can always solve the equation order by order in a radial expansion with appropriate boundary conditions. However, this does not constitute a proof of the existence of a globally well defined solution. Indeed, we showed in Section 5.3 that a global solution fails to exist for particular pp-waves, albeit in the case of $\mu\ell >1$. Similar phenomenon has also been noticed in \cite{adiherif} in their study of the ordinary Witten spinor equation in pure AdS gravities in diverse dimensions. The role of spin structures and whether they can be extended to the bulk is another global issue which requires careful study \cite{adiherif}.

Our focus has been primarily on $\mu\ell =1$ theory with standard
Brown-Henneaux boundary conditions. If we allow the weak version of
these boundary conditions given in \eq{cbc}, the positivity condition
for $Q_{ADT}[\xi^{(+)}]$ remains the same as in \eq{ct1} but
$Q_{ADT}[\xi^{(-)}]$ no longer vanishes, and is given by
\be
\mu\ell=1:\qquad Q_{ADT} [\xi^{(-)}]
=  \frac1{16\sqrt{2} \pi G} \int d\phi \, e^{-\rho} \,
\xi^{(-)-} \tf_{--} \ .
\label{cpe}
\ee
Surprisingly, there is no Witten-Nester identity available for this charge, and consequently whether it is positive or not depends on the outcome of a direct evaluation of the above integral, as discussed in Section 3.3. For the exact pp-wave solution \eq{sss3}, the result for $Q_{ADT}[\xi^{(-)}]$ is given by \eq{cpe} with $\tf_{--}=4f_1(u)$, which suggests that one can always choose $f_1$ such that this charge is negative. This is consistent with the fact that there exist negative energy linearized solutions which satisfy the weak Brown-Henneaux boundary conditions \cite{Grumiller1}. Therefore, these boundary conditions must be ruled out in the chiral theory.

In the case of $\mu\ell>1$, the ADT charges are given in \eq{05} and \eq{cases} (with standard Brown-Henneaux boundary conditions \eq{sbc}) and we have the bounds
\be
\mu\ell >1:\qquad \ \ Q_{ADT}[\xi^{(\pm)}]  \ge
- \frac{1}{8\pi G \mu^3(\mu\pm m)}
\int_\Sigma da\, X_{\mu\nu}\, u^\mu v^{(\pm)\nu}\ ,
\label{mpc6}
\ee
where $X_{\mu\nu}$ is defined in \eq{xdef}, and $v^{(\pm)\nu}$ are bilinear in Witten spinors which approach the left or right Killing spinors asymptotically. As we saw earlier there is no reason for these expressions to be positive or vanishing with the exception of Type $N$ spacetimes, and Type $I$ spacetimes with $\alpha=\beta$ (see Tables above). In particular the pp-waves are of Type $N$, and in this case the ADT charges can be directly computed, giving the results
\bea
Q_{ADT}[\xi^{(+)}] &=& -\frac{1}{8\sqrt{2} \pi G} \int d\phi \,
\left( 1 + \frac{m}{\mu} \right) e^{-\rho} \xi^{(+)+} > 0\ ,
\w2
Q_{ADT}[\xi^{(-)}]  &=&
\frac{1}{8\sqrt{2} \pi G} \int d\phi
\left( 1 - \frac{m}{\mu} \right) e^{-\rho} \, \xi^{(-)-}
\left( 2f_3(u) + 1 \right) \ge 0\ ,
\label{b22}
\eea
where $f_3(u)$ is the function occurring in the pp-wave solution \eq{sss1}, and the bound  \eq{b22}, which follows from \eq{mpc6}, holds provided that a globally well defined solution of the generalized Witten spinor equation exists for this solution. This bound seems
puzzling at first sight because one may consider a function $f_3$ for which the ADT energy is negative. However, for any such choice of $f_3$  a regular Witten spinor must fail to exist. Indeed, as we saw in Section 5.3, the exact solution of the generalized Witten equation is not regular for constant $f_3$ that gives negative ADT energy. Of course, we do not know if the Type $N$ solutions are the only ones that are asymptotically AdS. Nonetheless, the bound \eq{b22} suggests that the energy may be positive for all Type $N$ spacetimes, in a manner similar to the case of $\mu\ell=1$. On the other hand, it is known that (see, for example, \cite{LSS}) there exists a linearized solution of CTMG which has negative energy helicity 2 excitation for $\mu\ell>1$. We do not know if this result survives necessarily at the nonperturbative level. If we assume that it does, then we must investigate whether the requirement that the generalized Witten equation admits a globally well defined solution is too restrictive in determining the full space of solutions with standard Brown-Henneaux boundary conditions.

\section*{Acknowledgements}

We are indebted to Andy Strominger and Stanley Deser for discussions and correspondence which motivated this work. We thank David Chow, Gary Gibbons, Daniel Grumiller, Chris Pope and Massimo Porrati for helpful discussions. The research of E.S. is supported in part by NSF grant PHY-0555575.

\newpage

\begin{appendix}


\section{Notations and Conventions} \label{app:conv}


In our conventions  $\eta_{ab}={\rm diag} (-1,+1,+1)$. The world
indices $\mu$ and the tangent space indices $a$ are split as
\bea
&& \mu = (\tau, \phi, \rho)\ ,
\nn\w2
&& a=(0,1,2)=(+,-,2)= (0, i)\ , \quad i=1,2\ .
\eea
The $\pm$ labels are reserved to flat indices only throughout the
paper. We define the light-cone indices in the local Lorentz frame as
\be
v_\pm = \frac{1}{\sqrt{2}} \left( \pm v_0 + v_1 \right)\ ,  \qquad
v^\pm = \frac{1}{\sqrt{2}} \left( \pm v^0 + v^1 \right) = v_\mp\ ,
\label{a1}
\ee
and the coordinates as
\be
u = \frac{1}{\sqrt{2}} \ell \, (-\tau+\phi)\ , \qquad
v = \frac{1}{\sqrt{2}} \ell \, (\tau+\phi)\ .
\ee
The Clifford algebra is
\be
\{\c^a,\c^b\}=2\eta^{ab}\ ,
\ee
and we use the representation
\be
\c_0=i\sigma_2 \ , \quad \c_1=\sigma_1\ ,\quad  \c_2=\sigma_3 \ .
\ee
Furthermore, we use the conventions
\be
\c^{abc}= \epsilon^{abc}\ , \quad \varepsilon_{\mu\nu\rho}=
e \epsilon_{\mu\nu\rho}\ ,\quad
\varepsilon^{\mu\nu\rho}= e^{-1}\epsilon^{\mu\nu\rho}\ ,
\ee
where $e = {\rm det}\ e_\mu^a$ and the $\epsilon$-tensors are constant.
The chiral projected $\c$-matrices $\c^\pm$ and $\c_\pm$ are defined as in
\eq{a1}.
The Dirac conjugate of a spinor $\psi$ is defined as
\be
\bar{\psi} = \psi^\dagger \gamma^0\ .
\label{dconj}
\ee


\section{The Noether Current and Superpotential } \label{app:silva}


Following \cite{silva}, we summarize the main points of how to construct
a Noether current associated with local symmetries.
In Section 3 we apply this procedure to CTMG.

Consider the Lagrangian $\cL(\phi,\partial\phi)$ that possesses a local symmetry with gauge parameters $\xi^a(x)$, and therefore satisfying $\delta \cL = \partial_\mu S^\mu_\xi$. Noether's second theorem then implies the existence of the on-shell conserved current $J^\mu_\xi$ such that
\be
\partial_\mu J^\mu_\xi = \delta_\xi \phi \frac{\delta \cL}{\delta\phi} = 0\ ,
\label{s1}
\ee
where $\frac{\delta {\cal L}}{\delta \phi} = 0$ is the Euler-Lagrange
equation of $\phi$, and
\be
 J^\mu_\xi := S^\mu_\xi -\delta_\xi\phi
\frac{\partial \cL}{\partial\partial_\mu\phi}\ .
\ee
Let us parametrize the variation of the field as follows
\be
\delta \phi = \xi^a \Delta_a(\phi) + \partial_\nu \xi^a \Delta_a^\nu(\phi)\ .
\ee
It is clear that the surface term $S_\xi^\mu$ is not uniquely defined since any transformation $S_\xi^\mu \rightarrow S_\xi^\mu + \partial_\nu S_\xi^{\mu\nu}$ such that $S_\xi^{\mu\nu}= -S_\xi^{\nu\mu}$ does not modify $\partial_\mu S_\xi^\mu$. Let us make the a priori arbitrary choice such that \cite{silva}
\be
S^\mu_\xi = \xi^a \Sigma^\mu_a(\phi) +\partial_\nu \xi^a \Sigma^{\mu\nu}_a (\phi)\ .
\ee
Using the abelian restriction $\xi^a(x) := \epsilon(x) \xi^a_0(x)$ where $\epsilon(x)$ is an arbitrary function and $\xi^a_0(x)$ is fixed but spacetime dependent, after some manipulations one finds from \eq{s1} that \cite{silva}
\be
J^\mu_\xi = \partial_\nu U^{\mu\nu}_\xi+ \xi^a \Delta^\mu_a \frac{\delta\cL}{\delta\phi}\ ,
\label{silva}
\ee
where $U^{\mu\nu}_{\xi^0} := \xi^a_0 U^{\mu\nu}_a =-U^{\nu\mu}_{\xi_0}$ is called the superpotential,  with
\be
U^{\mu\nu}_a := \Sigma^{\mu\nu}_a -\Delta^\nu_a \frac{\partial\cL}{\partial\partial_\mu\phi}\ ,
\ee
and the subscript $0$ has been dropped everywhere. The conserved charge is given by the integral of the superpotential at spatial infinity as
\be
Q(\xi)= \int_{\partial \Sigma} U^{\mu\nu}_\xi d\Sigma_{\mu\nu}\ .
\ee
A key question is how to choose $\Sigma^{\mu\nu}_a$ such that this charge generates the appropriate transformations through the Poisson bracket, as has been emphasized in \cite{Henneaux:1999ct}. Here we shall follow the proposal of \cite{silva}, according to which $U_\xi^{\mu\nu}$ is chosen such that the variation of the Noether current is localizable in the sense that
\be
\delta J^\mu_\xi = \delta\phi \frac{\delta W^\mu_\xi}{\delta\phi}\ ,
\ee
where
\be
W^\mu_\xi := \xi^a \Delta^\mu_a \frac{\delta\cL}{\delta\phi}\ .
\label{w}
\ee
Note, in particular, that no $\partial_\mu \delta\phi$ terms are present as a requirement of this prescription. This proposal is motivated by the Hamiltonian approach of Regge and Teitelboim \cite{Regge:1974zd}. Furthermore this definition is independent of the choice of surface terms added to the Lagrangian since it depends only on the field equations through $W^\mu_\xi$ as in  \eq{w}, and it has been shown to produce the appropriate conserved charges in many examples \cite{silva} including supergravity \cite{Henneaux:1999ct}. We shall use this prescription for the construction of the Noether supercurrent and superpotential, and show that the associated conserved charges have the desired property for the CTMG.


\section{The Abbott-Deser-Tekin Charges for TMG in Arbitrary Background} \label{app:adt}


The Abbott-Deser procedure \cite{Abbott:1981ff} for defining the
conserved quantities in
asymptotically AdS spacetimes was generalized to higher derivative
theories, and in particular to TMG by Deser and Tekin \cite{Deser:2003vh}.
This was further generalized by Bouchareb and Cl\'ement \cite{Bouchareb:2007yx}
to TMG with arbitrary backgrounds. We shall follow \cite{Bouchareb:2007yx}
to summarize these results here.

Given a solution ${\bar g}_{\mu\nu}$ of the field equation (\ref{fe1})
$\cE_{\mu\nu} \equiv {\cal G}_{\mu\nu} + \mu^{-1} C_{\mu\nu} = 0$,
we can write
\be
g_{\mu\nu}= {\bar g}_{\mu\nu} +h_{\mu\nu}\ ,
\ee
where $h_{\mu\nu}$ represents the deviation (not necessarily small) from
the background solution. Then, it follows from the Bianchi identity
$\nabla_\nu \cE^{\mu\nu}=0$ that linearized tensor $\delta \cE_{\mu\nu}$
is conserved as follows
\be
\bar\nabla_\nu \delta \cE^{\mu\nu}=0\ .
\label{b}
\ee
If the background admits a Killing vector $\xi^\mu$, then the current $ \delta\cE^{\mu\nu} \xi_\nu$ is covariantly conserved. It follows that there exists an antisymmetric tensor field $\cF^{\mu\nu}$ such that
\be
\delta\cE^{\mu\nu} \xi_\nu = \bar\nabla_\nu \cF^{\mu\nu}\ ,
\ee
and the conserved charge is defined as
\be
Q(\xi) = \frac1{8\pi G} \int_\Sigma \delta\cE^{\mu\nu}
\xi_\nu \, d\Sigma_\mu
= \frac{1}{8\pi G} \int_{\partial \Sigma} \cF^{\mu\nu}
d\Sigma_{\mu\nu}\ . \
\ee
For the TMG model we are studying, it is found
in \cite{Bouchareb:2007yx} that
\be
\xi_\nu \delta\cG^{\mu\nu}
= \bar{\nabla}_\nu \cF^{\mu\nu}_E (\xi)
- \xi^\nu \bar{\cG}^{\mu\lambda} h_{\lambda\nu}
+ \frac12 \xi^\mu \bar{\cG}^{\lambda\rho} h_{\lambda\rho}
- \frac12\xi^\nu \bar{\cG}^\mu{}_\nu h\ ,
\label{c1}
\ee
where $h\equiv \bar{g}^{\mu\nu} h_{\mu\nu}$ and
\bea
\cF^{\mu\nu}_E (\xi)
&=& \frac12 \big(\xi^\nu \bar\nabla_\lambda h^{\lambda\mu}
-\xi^\mu \bar\nabla_\lambda h^{\lambda\nu}
+\xi_\lambda \bar\nabla^\mu h^{\lambda\nu}
-\xi_\lambda \bar\nabla^\nu h^{\lambda\mu} +\xi^\mu \bar\nabla^\nu h
-\xi^\nu \bar\nabla^\mu h
\nn\w2
&& +h^{\nu\lambda} \bar\nabla_\lambda \xi^\mu
-h^{\mu\lambda} \bar\nabla_\lambda \xi^\nu
+ h \bar\nabla^\mu \xi^\nu\big)\ .
\label{fe}
\eea
From the Cotton tensor, it is found that \cite{Bouchareb:2007yx}
\be
\xi_\nu \delta C^{\mu\nu}
= \bar\nabla_\lambda \cF^{\mu\lambda}_C (\xi)
- \xi^\nu \bar{C}^{\mu\lambda} h_{\lambda\nu}
+\frac12 \xi^\mu \bar{C}^{\lambda\rho} h_{\lambda\rho}
-\frac12\xi^\nu \bar{C}^\mu{}_\nu h\ ,
\label{c2}
\ee
and
\bea
\cF^{\mu\nu}_C (\xi) &=& \cF^{\mu\nu}_E (\Xi)
+ \bar{\varepsilon}^{\mu\nu\rho} \xi_\lambda
\left( \delta \cG^\lambda{}_\rho -\frac12 \delta^\lambda_\rho \delta\cG\right)
\nn\w2
&& + \frac12 \bar{\varepsilon}^{\mu\nu\rho} \left[
\xi_\rho  \bar{G}^\sigma{}_\lambda h^\lambda{}_\sigma
+\frac12 \left(\xi_\sigma \bar{G}^\sigma{}_\rho
+ \frac12 \xi_\rho \bar{R} \right)h \right]\ ,
\label{fc}
\eea
where $\delta \cG^\mu{}_\nu = \delta G^\mu{}_\nu$ denotes the linear
in $h$ deviation of $\cG^\mu{}_\nu$ from its background value
$\bar\cG^\mu{}_\nu$, and
\be
\Xi^\mu \equiv \frac12 \bar\varepsilon^{\mu\nu\rho} \bar\nabla_\nu \xi_\rho\ .
\label{Xi}
\ee
Summing up the contributions \eq{c1} and \eq{c2}, the $\bar{\cG} h$
and $\bar{C} h$ terms sum up to give the field equation $\cE_{\mu\nu}=0$
thereby vanishing, giving the Bouchareb and Cl\'ement result
\cite{Bouchareb:2007yx}
\be
Q_{BC}[\xi] = \frac1{8\pi G} \int_{\partial \Sigma}
\left( \cF^{\mu\nu}_E(\xi) +\frac1{\mu} \cF^{\mu\nu}_C (\xi)\right)
d\Sigma_{\mu\nu}\ ,
\ee
with $\cF^{\mu\nu}_E(\xi)$ and  $\cF^{\mu\nu}_C(\xi)$ given in
\eq{fe} and \eq{fc}, respectively
\footnote{In comparing this result with that of \cite{Deser:2003vh},
observe that the last three terms in \eq{fc} vanish for $AdS_3$
background, and that $
\bar{\varepsilon}^{\mu\nu\rho} \left( \delta{\cal G}_{\rho\sigma}
- \frac{1}{2} \bar{g}_{\rho\sigma} \delta{\cal G} \right) \xi^\sigma
= \frac12\Big(
 \bar{\varepsilon}^{\mu\nu\rho} \delta{\cal G}^\sigma{}_\rho \xi_\sigma
+ \bar\varepsilon^{\sigma\nu\rho} \delta{\cal G}^\mu{}_\rho \xi_\sigma
+ \bar\varepsilon^{\mu\sigma\rho} \delta{\cal G}^\nu{}_\rho \xi_\sigma\Big)
$, as can readily be seen from the identity
$\bar\varepsilon^{[\sigma\nu\rho} \delta{\cal G}^{\mu]}{}_\rho = 0$.}.
In general $\Xi^\mu$ is not a Killing vector. However, in the case of $AdS_3$ background they are Killing vectors, and in fact $\Xi^\mu=m\xi^\mu$ when $\xi^\mu$ can be written in terms of Killing spinors as in \eq{k1}.  Furthermore, the last three terms in \eq{fc} sum up to zero in $AdS_3$ background, and thus for a Killing vector satisfying $\Xi^\mu= m\xi^\mu$ we obtain the simplified result obtained first by Deser and Tekin \cite{Deser:2003vh}
\bea
Q_{ADT}[\xi]
&=& \frac1{8\pi G} \int_{\partial \Sigma}
\left( \cF^{\mu\nu}_E(\xi) + \frac{1}{\mu} \cF^{\mu\nu}_E(\Xi)
+ \frac1{\mu} f^{\mu\nu}_C (\xi)\right) d\Sigma_{\mu\nu}
\label{dtf}\w2
&=& \frac1{8\pi G} \int_{\partial \Sigma}
\left[ \left( 1 + \frac{m}{\mu} \right) \cF^{\mu\nu}_E(\xi)
+ \frac1{\mu} f^{\mu\nu}_C (\xi)\right] d\Sigma_{\mu\nu}\ ,
\label{dt}
\eea
where
\bea
f^{\mu\nu}_C (\xi) &=& \bar{\varepsilon}^{\mu\nu\rho}
\left( \delta \cG_{\rho\sigma}
-\frac12 \bar{g}_{\rho\sigma} \delta\cG\right)\xi^\sigma\ ,
\label{line1}
\w2
&=& \bar{\varepsilon}^{\mu\nu\rho}\delta \cG_{\rho\sigma} \xi^\sigma \ ,
\label{line2}
\eea
and in obtaining the second line we have used the consequence of the field equation \eq{fe1} giving ${\cal G} \equiv {\cal G}^\mu{}_\mu = 0$, and therefore the second term in \eq{line1} vanishes. Note also that $\delta \cG^\lambda{}_\nu ={\bar  g}^{\lambda\mu}\, \delta \cG_{\mu\nu}$, where we have restored the bar notation for background metric for clarity, for backgrounds with ${\bar \cG}_{\mu\nu}=0$.

For a Killing vector $\tilde{\xi}^\mu$ satisfying $\tilde{\Xi}^\mu = - m \tilde{\xi}^\mu$ (which is the case when \eq{k2} holds), the ADT charge becomes
\be
Q_{ADT}[\tilde{\xi}] = \frac{1}{8\pi G} \int_{\partial \Sigma}
\left[ \left( 1 - \frac{m}{\mu} \right)
{\cal F}_E^{\mu\nu}(\tilde{\xi})
+ \frac{1}{\mu} f_C^{\mu\nu}(\tilde{\xi}) \right] d\Sigma_{\mu\nu}\ .
\label{E+J}
\ee

Finally, we justify the overall normalization of the energy in \eq{dtf}
as follows. Let us consider a coupling to a scalar field
\be
{\cal L} = \frac{1}{16\pi G} \sqrt{-g} ( R + 2m^2 )
+ \frac{1}{16\pi G\mu} {\cal L}_{CS}
- \frac{1}{2} \sqrt{-g} g^{\mu\nu} \partial_\mu \phi
\partial_\nu \phi\ .
\ee
Here we have taken the coefficient of the Einstein term to be the
canonical one as in \eq{1.1}. The field equation for the gravitational
field is
\be
\frac{1}{8\pi G} \left({\cal G}_{\mu\nu}
+ \frac{1}{\mu} C_{\mu\nu}\right) = T_{\mu\nu}^{\rm (matter)}\ ,
\ee
where
\be
T_{\mu\nu}^{\rm (matter)}
= \partial_\mu \phi \partial_\nu \phi
- \frac{1}{2} g_{\mu\nu} (\partial \phi)^2
\ee
is the standard energy-momentum tensor of the scalar field.
Splitting the left-hand side into a part linear in $h_{\mu\nu}$
and a higher order part following \cite{Deser:2003vh} we obtain
\be
\frac{1}{8\pi G} \left( \delta{\cal G}_{\mu\nu}
+ \frac{1}{\mu} \delta C_{\mu\nu} \right)
= T_{\mu\nu}^{\rm (matter)} + T_{\mu\nu}^{\rm (gravity)}
\equiv T_{\mu\nu}\ .
\ee
Energy is defined such that the matter part becomes the standard
scalar energy and is given by
\bea
E &=& \int_\Sigma T^\mu{}_\nu \, \xi^\nu d\Sigma_\mu
\nn\w2
&=& \frac{1}{8\pi G} \int_\Sigma \left( \delta{\cal G}^\mu{}_\nu
+ \frac{1}{\mu} \delta C^\mu{}_\nu \right) \xi^\nu d\Sigma_\mu
\nn\w2
&=& \frac{1}{8\pi G} \int_{\partial\Sigma} \left[
\cF_E^{\mu\nu}(\xi) + \frac{1}{\mu} \cF_E^{\mu\nu}(\Xi)
+ \frac{1}{\mu} f_C^{\mu\nu}(\xi) \right] d\Sigma_{\mu\nu}\ ,
\eea
in agreement with \eq{dtf}.


\section{$AdS_3$ Background, Killing Spinors and Killing Vectors} \label{app:killing}


The metric of the background AdS spacetime is
\be
d\bar{s}^2 = dx^\mu dx^\nu \bar{g}_{\mu\nu}
= \ell^2 \left[
- \cosh^2\rho \, d\tau^2 + \sinh^2\rho \, d\phi^2 + d\rho^2 \right]\ .
\label{adsmetric}
\ee
We use the world indices $\mu = \tau, \phi, \rho$ and the local Lorentz
indices $a = 0, 1, 2$. The dreibein is chosen as
$\bar{e}_\tau{}^0 = \ell \cosh\rho$, $\bar{e}_\phi{}^1 = \ell \sinh\rho$
and $\bar{e}_\rho{}^2 = \ell$. The non-vanishing components of the spin
connection are $\bar{\omega}_\tau{}^{02} = \sinh\rho$,
$\bar{\omega}_\phi{}^{12} = \cosh\rho$. The background value of
\be
\hat{\nabla}^L_\mu
= \partial_\mu + \frac{1}{4} \omega_\mu{}^{ab} \gamma_{ab}
- \frac{1}{2}m \gamma_\mu
\ee
is given by
\bea
\bar{\hat{\nabla}}_\tau^L
&=& \partial_\tau - \frac{1}{4} e^\rho \gamma_0 (1-\gamma_2)
- \frac{1}{4} e^{-\rho} \gamma_0 (1+\gamma_2)\ ,
\nn\w2
\bar{\hat{\nabla}}_\phi^L
&=& \partial_\phi - \frac{1}{4} e^\rho \gamma_0 (1-\gamma_2)
- \frac{1}{4} e^{-\rho} \gamma_0 (1+\gamma_2)\ ,
\nn\w2
\bar{\hat{\nabla}}_\rho^L
&=& \partial_\rho - \frac{1}{2} \gamma_2\ .
\eea
For some purposes it is more convenient to express the AdS metric as
\be
d\bar{s}^2 = - \left( 1 + m^2r^2 \right) dt^2
+ r^2 d\phi^2
+ \frac{dr^2}{1 + m^2r^2}\ ,
\label{adsmetric2}
\ee
which is related to \eq{adsmetric} by a change of coordinates
$t = \ell\tau$, $r = \ell \sinh\rho$ with $m=\ell^{-1}$.

There are two independent Killing spinors satisfying
$\bar{\hat{\nabla}}^L_\mu \epsilon_K = 0$
in the AdS background \cite{gps}
\bea
\epsilon_K^1 &=& \left[
e^{\frac{1}{2}\rho} \cos\ft12(\tau+\phi)
+ \gamma_0 e^{-\frac{1}{2}\rho} \sin\ft12(\tau+\phi)
\right] \eta_+^1\ ,
\nn\w2
\epsilon_K^2 &=& \left[
e^{\frac{1}{2}\rho} \sin\ft12(\tau+\phi)
- \gamma_0 e^{-\frac{1}{2}\rho} \cos\ft12(\tau+\phi)
\right] \eta_+^2\ ,
\label{kspinor}
\eea
where $\eta_+^A$ ($A=1,2$) are arbitrary constant Majorana spinors
satisfying $\gamma_2 \eta_+^A = \eta_+^A$.
We can construct three Killing vectors from these (commuting)
Killing spinors as
\be
\xi^{AB\,\mu} = \bar{\epsilon}_K^{A} \bar\gamma^\mu
\epsilon_K^{B} = \xi^{BA\,\mu} \qquad (A,B=1,2)\ .
\label{k1}
\ee
The explicit forms of $\xi^{AB}= \xi^{AB\mu} \partial_\mu$ are
\bea
\xi^{11} &=& -2 \ell^{-1} \left( K_0 + K_1 \right)\ ,
\nn\w2
\xi^{22} &=& -2 \ell^{-1} \left( K_0 - K_1 \right)\ ,
\nn\w2
\xi^{12} &=& -2 \ell^{-1} K_2\ ,
\eea
where the normalization of $\eta^A_+$ has been chosen as
$\bar{\eta}_+^A \gamma^0 \eta_+^B = -1$, and $K_a^\mu$ are given by
\bea
K_0 &=& \ft12 \left(\ptau+\pphi\right)\ ,\nn\w2
K_1 &=& \ft12 \cos(\tau+\phi)\left(
\tanh\rho \ptau +\coth\rho\pphi \right)
+ \ft12 \sin(\tau+\phi)\prho\ , \nn\w2
K_2 &=& \ft12 \sin(\tau+\phi)\left(
\tanh\rho \ptau +\coth\rho\pphi\right)
- \ft12 \cos (\tau+\phi)\prho\ .
\label{lkv}
\eea
Since the vectors $K_a$ satisfy
\be
K_a^\mu K_{b\mu} = \frac{1}{4} \ell^2 \, \eta_{ab}\ ,
\ee
$K_0 \pm K_1$ are null. Furthermore, $K_a=K_a^\mu \partial_\mu$ obey
the algebra $[K_a,K_b]= -\epsilon_{ab}{}^c K_c$.
The three Killing vectors $\xi^{AB\,\mu}$ are generators of $SO(1,2)_L$ in
$SO(2,2) = SO(1,2)_L \times SO(1,2)_R$.

Other three Killing vectors, which are generators of $SO(1,2)_R$, are
constructed from Killing spinors satisfying
$\bar{\hat{\nabla}}^R_\mu \epsilon_K = 0$, where
$\bar{\hat{\nabla}}^R_\mu = \bar{\nabla}_\mu
+ \frac{1}{2}m \bar{\gamma}_\mu$. These solutions are
\bea
\epsilon_K^{\dot{1}} &=& \left[
e^{\frac{1}{2}\rho} \cos\ft12(\tau-\phi)
- \gamma_0 e^{-\frac{1}{2}\rho} \sin\ft12(\tau-\phi)
\right] \eta_-^{\dot{1}}\ ,
\nn\w2
\epsilon_K^{\dot{2}} &=& \left[
e^{\frac{1}{2}\rho} \sin\ft12(\tau-\phi)
+ \gamma_0 e^{-\frac{1}{2}\rho} \cos\ft12(\tau-\phi)
\right] \eta_-^{\dot{2}}\ ,
\label{rc}
\eea
where $\eta_-^{\dot{A}}$ ($\dot{A}=\dot{1},\dot{2}$) are arbitrary
constant Majorana spinors satisfying
$\gamma_2 \eta_-^{\dot{A}} = - \eta_-^{\dot{A}}$,
and the three Killing vectors can be written as
\be
\xi^{\dot{A}\dot{B}\,\mu}
= \bar{\epsilon}_K^{\dot A} \bar\gamma^\mu \epsilon_K^{\dot B}
= \xi^{\dot{B}\dot{A}\,\mu}
\qquad (\dot{A}, \dot{B} = \dot{1}, \dot{2})\ .
\label{k2}
\ee
The explicit forms of
$\xi^{\dot A\dot B}= \xi^{\dot A\dot B\mu} \partial_\mu$ are
\bea
\xi^{\dot 1\dot 1} &=& -2 \ell^{-1} \left( J_0 + J_1 \right)\ ,
\nn\w2
\xi^{\dot 2\dot 2} &=& -2 \ell^{-1} \left( J_0 - J_1 \right)\ ,
\nn\w2
\xi^{\dot 1\dot 2} &=& 2 \ell^{-1} J_2\ ,
\eea
where $J_a^\mu$ are given by
\bea
J_0 &=& \ft12  \left(\ptau-\pphi\right)\ ,
\nn\w2
J_1 &=& \ft12 \cos(\tau-\phi)\left( \tanh\rho \ptau
-\coth\rho\pphi\right) +\ft12 \sin(\tau-\phi)\prho\ ,
\nn\w2
J_2 &=& -\ft12 \sin(\tau-\phi)\left( \tanh\rho \ptau
-\coth\rho\pphi\right) +\ft12 \cos (\tau-\phi)\prho\ ,
\label{rkv}
\eea
and $J_a=J^\mu_a \partial_\mu$ obey the algebra
$[J_a,J_b]= \epsilon_{ab}{}^c J_c$.

Finally, we use the abbreviated notation $\xi^{(+)}$ and $\xi^{(-)}$ for
$\xi^{AB}$ and $\xi^{\dot A \dot B}$, respectively, with components
given by $\xi^{(\pm)} = \xi^{(\pm)+}\partial_+ + \xi^{(\pm)-}\partial_-
+ \xi^{(\pm)2}\partial_2$.  Note in particular that
$-2\ell^{-1}K_0 = \ft12 ( \xi^{11} + \xi^{22} )$ and
$-2\ell^{-1}J_0 = \ft12 ( \xi^{\dot 1\dot 1} + \xi^{\dot 2 \dot 2} )$,
associated with $(E-mJ)$ and $(E+mJ)$, respectively, are given by
\bea
-2 \ell^{-1} K_0 &=& - \frac{1}{\sqrt{2}} \, \left(
e^\rho \partial_+ - e^{-\rho} \partial_- \right)\ ,
\nn\w2
-2 \ell^{-1} J_0 &=& \frac{1}{\sqrt{2}} \, \left(
e^\rho \partial_- - e^{-\rho} \partial_+ \right)\ .
\label{kn}
\eea

\end{appendix}

\end{document}